\documentclass[aps,prd,twocolumn,amsmath,amssymb,showpacs,footnote,reprint]{revtex4-1}

\usepackage{graphicx}
\usepackage{dcolumn}
\usepackage{bm}

\usepackage{epstopdf} 

\usepackage[colorlinks,hyperindex]{hyperref}  
\hypersetup
    {
        colorlinks,%
        citecolor=blue,%
        linkcolor=blue,%
        urlcolor=blue,%
    }

\usepackage{multirow} 
\allowdisplaybreaks[1] 

\begin{document}

\title{Renormalized stress tensor for massive fields \\
in Kerr-Newman spacetime}

\author{Andrei Belokogne}
\email{belokogne.andrei@gmail.com}

\author{Antoine Folacci}
\email{folacci@univ-corse.fr}

\affiliation{Equipe Physique
Th\'eorique, Projet COMPA, \\ SPE, UMR 6134 du CNRS
et de l'Universit\'e de Corse,\\
Universit\'e de Corse, BP 52, F-20250 Corte,
France}

\date{\today}

\begin{abstract}

In a four-dimensional spacetime, the DeWitt-Schwinger expansion of the effective action associated with a massive quantum field reduces, after renormalization and in the large mass limit, to a single term constructed from the purely geometrical Gilkey-DeWitt coefficient $a_3$ and its metric variation provides a good analytical approximation for the renormalized stress-energy tensor of the quantum field. Here, from the general expression of this tensor, we obtain analytically the renormalized stress-energy tensors of the massive scalar field, the massive Dirac field and the Proca field in Kerr-Newman spacetime. It should be noted that, even if, at first sight, the expressions obtained are complicated, their structure is in fact rather simple, involving naturally spacetime coordinates as well as the mass $M$, the charge $Q$ and the rotation parameter $a$ of the Kerr-Newman black hole and permitting us to recover rapidly the results already existing in the literature for the Schwarzschild, Reissner-Nordstr\"om and Kerr black holes (and to correct them in the latter case). In the absence of exact results in Kerr-Newman spacetime, our approximate renormalized stress-energy tensors could be very helpful, in particular to study the backreaction of massive quantum fields on this spacetime or on its quasinormal modes.

\end{abstract}

\pacs{04.62.+v, 04.70.Dy}

\maketitle

\section{Introduction}

Quantum field theory in curved spacetime (for reviews and monographs on this
subject, see Refs.~\cite{DeWitt:1975ys,Birrell:1982ix,Fulling:1989nb,Wald:1995yp,Ford:2005qz,Parker:2009uva,Hollands:2014eia}) is a semiclassical approximation of quantum gravity which, by treating classically the spacetime metric $g_{\mu \nu}$ and considering from a quantum point of view all the other fields (including the graviton field to at least one-loop order for reasons of consistency), avoids the difficulties due to the nonrenormalizability of quantum gravity and provides a framework which permits us to study the low-energy consequences of a hypothetical ``theory of everything". It should be recalled that this approach allowed theoretical physicists to obtain fascinating results concerning early universe cosmology and quantum black hole physics (see Refs.~\cite{DeWitt:1975ys,Birrell:1982ix,Fulling:1989nb,Wald:1995yp,Ford:2005qz,Parker:2009uva,Hollands:2014eia} and references therein) and, in particular, led to the discovery of particle creation in expanding universes by Parker \cite{Parker:1969au} and of black hole radiance by Hawking \cite{Hawking:1974sw}. Furthermore, this approach also provides the natural theoretical framework to analyze the cosmic microwave background (CMB) observations made in recent years.

In quantum field theory in curved spacetime, it is conjectured that the backreaction of a
quantum field on the spacetime geometry is governed by the semiclassical Einstein equations
\begin{equation}\label{SCEinsteinEq}
G_{\mu \nu}=8 \pi \langle T_{\mu \nu}  \rangle_\mathrm{ren}.
\end{equation}
(In this article, we use the geometrical conventions of
Misner, Thorne and Wheeler \cite{Misner:1974qy} and we adopt units such that $\hbar = c = G_N = 1$.) In Eq.~(\ref{SCEinsteinEq}), $G_{\mu \nu}$ is the Einstein tensor $R_{\mu
\nu}-\frac{1}{2}g_{\mu \nu}R+ \Lambda g_{\mu \nu}$ or some higher-order generalization
of this geometrical tensor while $\langle
T_{\mu \nu}  \rangle_\mathrm{ren}$ is the renormalized stress-energy tensor (RSET) of the quantum field or, more precisely, the renormalized expectation value of the stress-energy tensor operator associated with the quantum field.

The semiclassical Einstein equations (\ref{SCEinsteinEq}) have permitted Starobinsky to show that, after the Planck era, quantum effects lead to an inflationary universe, i.e., a universe with an exponentially expanding de Sitter phase \cite{Starobinsky:1980te}. They have been also used by several authors to analyze the dynamics of evaporating black holes due to Hawking radiation (see, e.g., Refs.~\cite{Bardeen:1981zz,Hiscock:1980ze} for pioneering work on this topic) and, more recently, they have been considered to explain the acceleration of the expansion of the universe (see, e.g., Ref.~\cite{Parker:2003as}). But unfortunately, and despite the impressive successes mentioned above, the backreaction problem is in general difficult to tackle, not only because the semiclassical Einstein equations (\ref{SCEinsteinEq}) are a set of coupled nonlinear hyperbolic partial differential equations but also because it is in general difficult to define the right-hand side of these equations, i.e., to construct the RSET. Indeed, the expectation value of the stress-energy tensor operator is formally infinite and it is necessary to regularize it, i.e., to extract from this formally infinite quantity a meaningful finite part, and then to renormalize all the coupling constants appearing in the problem in order to remove the remaining infinite part. Much work has been done since the mid-1970s in connection with this subject (see Refs.~\cite{DeWitt:1975ys,Birrell:1982ix,Fulling:1989nb,Wald:1995yp,Ford:2005qz,Parker:2009uva,Hollands:2014eia} and references therein) and, currently, we have at our disposal some powerful procedures (such as the adiabatic regularization method, the dimensional regularization method, the $\zeta$-function approach, the point-splitting method...) permitting us to construct theoretically and without ambiguity the RSET.

It is however important to note that, in four-dimensional gravitational backgrounds, except if we work under very strong hypotheses (e.g., if we consider field theories in maximally symmetric spacetimes or if we study massless or conformally invariant field theories on very particular spacetimes), it is in general impossible to obtain an analytical expression for the RSET. In fact, in most cases, it is even impossible to construct, from a practical point of view, the RSET and, when this is possible, it is necessary in many cases to perform a numerical analysis in order to extract the physical content of the RSET and, of course, this does not simplify the backreaction problem. So, it is interesting to note that various approaches have been developed which permit us to deal with situations presenting a ``lower degree of symmetry" and to construct, in this context, accurate analytical approximations of the RSET (see, e.g., Refs.~\cite{Page:1982fm,Brown:1985ri,Brown:1986jy} for the theoretical bases of the ``Brown-Ottewill-Page approximation" which is limited to static Einstein spacetimes or Refs.~\cite{DeWitt65,DeWitt:2003pm,Avramidi:1986mj,Avramidi:2000bm} for the theoretical bases of the ``DeWitt-Schwinger approximation" which can be used in an arbitrary spacetime but is limited to massive fields in the large mass limit \footnote{The expression DeWitt-Schwinger approximation could confuse the reader. Here, we intend the DeWitt-Schwinger approximation of the RSET. This approximation is constructed from the renormalized effective action (5), is given explicitly by (7), is only valid in the large mass limit and is purely geometrical. It is important to note that the DeWitt-Schwinger representation of the Feynman propagator permitted Christensen \cite{Christensen:1976vb,Christensen:1978yd} and other authors (see, e.g., Refs.~\cite{DeWitt:1975ys,Birrell:1982ix}) to develop the so-called DeWitt-Schwinger approach of the renormalization of the stress tensor. Such an approach provides, for all mass values, exact results (in general, they cannot be put in an analytical form) which take into account the geometry of the gravitational background as well as the quantum state of the field.}). From a theoretical point of view, such analytical approximations could be helpful to find self-consistent solutions to Eq.~(\ref{SCEinsteinEq}).

Here, we focus on the DeWitt-Schwinger approximation of the RSET. It is based on the DeWitt-Schwinger expansion of the effective action associated with a massive quantum field and, formally, it can be used only when the Compton length associated with the massive field is much less than a characteristic length constructed from the curvature of spacetime. In this context, an analytical expression for the RSET is directly obtained from the metric variation of the renormalized effective action associated with the massive field. Here, it is important to recall that the DeWitt-Schwinger expansion of the effective action is purely geometrical and to note that, as a consequence, the approximate RSET does not take into account the quantum state of the field. The literature concerning the DeWitt-Schwinger approximation and its applications is rich (see, e.g., Refs.~\cite{Frolov:1982fr,Frolov:1983ig,Frolov:1984ra,Anderson:1994hg,Taylor:1996yu,Hiscock:1997jt,Taylor:1999ic,
Matyjasek:1999an,Anderson:2000pg,Matyjasek:2000iy,Matyjasek:2006nu,FernandezPiedra:2007qq,Piedra:2009pf,Piedra:2010ur,Matyjasek:2011ub} for applications to black holes, wormholes and black strings and Refs.~\cite{Matyjasek:2013vwa,PhysRevD.89.084055} for a recent work concerning the Friedmann-Lema{\^\i}tre-Robertson-Walker universes) and it is important to note that the approximate RSETs obtained from the DeWitt-Schwinger expansion of the effective action seem to be in good agreement with exact results (see Ref.~\cite{Anderson:1994hg} where this has been discussed for the Reissner-Nordstr\"om black hole).

In this article, we intend to use the DeWitt-Schwinger approximation in order to construct the approximate RSET for massive fields in Kerr-Newman spacetime. Despite its physical importance, this particular spacetime has never been considered previously and this is certainly due to the complexity of the calculations involved. Let us remark that the problems treated in the references previously mentioned mainly concern spherically or cylindrically symmetric spacetimes and/or Einstein spacetimes (i.e., spacetimes satisfying $R_{\mu \nu}= \Lambda g_{\mu \nu}$) and that the Kerr-Newman spacetime does not belong to one of these categories. To facilitate our work, we shall write the general expression of the RSET in an irreducible form in order to reduce substantially the number of its terms and the size of calculations and we shall use \textit{Mathematica} packages allowing us to perform tensor algebra very efficiently. We shall then obtain, in Sec.~\ref{SectionII} (see also Appendix~\ref{KN_SET_coeff} for details), analytical expressions for the RSET of the massive scalar field, the massive Dirac field and the Proca field in Kerr-Newman spacetime. It should be noted that, even if, at first sight, the expressions we shall display are complicated, their structure is in fact rather simple, involving naturally spacetime coordinates as well as the mass $M$, the charge $Q$ and the rotation parameter $a$ of the Kerr-Newman black hole and permitting us to recover rapidly, in Sec.~\ref{SectionIII}, the results already existing in the literature for the Schwarzschild black hole \cite{Frolov:1982fr,FrolovNovikov:1998}, for the Reissner-Nordstr\"om black hole \cite{Anderson:1994hg,Matyjasek:1999an} and for the Kerr black hole \cite{Frolov:1983ig,Frolov:1984ra}. We shall moreover correct the result obtained for the Dirac field in the latter case. In Sec.~\ref{SectionIV}, we shall conclude by mentioning some possible applications of our results and by considering the shift in mass and angular momentum of the Kerr-Newman black hole dressed with a massive quantum field.

Before entering into the technical part of our work, it seems to us necessary to recall that, due to its fundamental importance in connection with the Hawking effect, the construction of the RSET in black hole spacetimes has often been discussed in the past 40 years and, in Schwarzschild and Reissner-Norstr\"om spacetimes (i.e., in static spherically symmetric black holes), we have now at our disposal exact expressions (which, however, must be analyzed numerically) for the RSET associated with various quantum fields and various vacuum states (see, e.g., Refs.~\cite{Candelas:1980zt,Fawcett:1983dk,Howard:1984qp,Howard:1985yg,Jensen:1988rh,Jensen:1991ef} for pioneering work concerning massless fields in Schwarzschild spacetime and Ref.~\cite{Anderson:1994hg} for results concerning both massless and massive fields in Schwarzschild and Reissner-Nordstr\"om spacetimes). For stationary axisymmetric black holes, the situation is less clear and much more complex. In Kerr spacetime (see, e.g., Refs.~\cite{Frolov:1989jh,Ottewill:2000qh,Ottewill:2000yr,Duffy:2005mz,Casals:2012es} and references therein), for massless fields, due both to problems linked with the quantization process in this spacetime \cite{Kay:1988mu,Frolov:1989jh} and to the complexity of the mode solutions of the wave equation, the RSET has been calculated only in specific locations and, for massive fields, to our knowledge, there exists no result concerning the RSET. In Kerr-Newman spacetime, the situation is even worse: we have found no results for this tensor in the literature.

\begin{table}[b!]
\caption{\label{tab:table1} The coefficients $c_i$ defining the renormalized effective action (\ref{EffAct_2}).}
\begin{ruledtabular}
\begin{tabular}{cccc}
      & \quad Scalar field  \quad & \quad Dirac field  \quad &   Proca field  \\
           & \quad $(s=0)$  \quad & \quad $(s=1/2)$  \quad &   $(s=1)$  \\
\hline
$c_1$ & $(1/2) \xi^2 - (1/5)\xi +1/56 $  & -3/280 & -27/280   \\
$c_2$ & $1/140 $  & 1/28 & 9/28 \\
\hline
$c_3$ & $-(\xi-1/6)^3 $  & 1/864 & -5/72 \\
$c_4$ & $(1/30) (\xi-1/6) $  & -1/180 & 31/60 \\
$c_5$ & $-8/945 $  & -25/756 & -52/63 \\
$c_6$ & $ 2/315 $  & 47/1260 & -19/105 \\
$c_7$ & $-(1/30) (\xi-1/6) $  & -7/1440 & -1/10 \\
$c_8$ & $1/1260 $  & 19/1260 & 61/140 \\
$c_9$ & $ 17/7560$  & 29/7560 & -67/2520 \\
$c_{10}$ & $-1/270 $  & -1/108 & 1/18
\end{tabular}
\end{ruledtabular}
\end{table}

We now describe the formalism we shall use in the following. We work in a four-dimensional spacetime $({\cal M}, g_{ab})$ without boundary and we consider more particularly the massive scalar field $\phi$ solution of the Klein-Gordon equation
\begin{equation}\label{EqOnde_s0}
(\Box -m^2 - \xi R )\phi=0,
\end{equation}
the massive spinor field $\psi$ solution of the Dirac equation
\begin{equation}\label{EqOnde_s1/2}
(\gamma^\mu \nabla_\mu + m)\psi=0
\end{equation}
and the massive vector field $A^\mu$ solution of the Proca equation
\begin{equation}\label{EqOnde_s1}
(g_{\mu \nu}\Box -m^2g_{\mu \nu} - \nabla_\mu \nabla_\nu - R_{\mu \nu} )A^\nu=0.
\end{equation}
In the wave equations (\ref{EqOnde_s0})-(\ref{EqOnde_s1}), $m$ denotes the mass of the fields, in the Klein-Gordon equation (\ref{EqOnde_s0}), $\xi$ is a
dimensionless factor which accounts for the possible coupling
between the scalar field and the gravitational background and in the Dirac equation (\ref{EqOnde_s1/2}), $\gamma^\mu$ denote the usual Dirac matrices. We recall that, after renormalization and in the large mass limit, the DeWitt-Schwinger expansion of the effective action associated with the massive scalar field, the massive Dirac field and the Proca field can be constructed from the Gilkey-DeWitt coefficient $a_3$ and reduces to \cite{DeWitt65,DeWitt:2003pm,Avramidi:1986mj,Avramidi:2000bm}
\begin{widetext}
\begin{eqnarray} \label{EffAct_2}
& & W_{\mathrm{ren}}= \frac{1}{192 \pi^2 m^2}\int_{\cal M} d^4
x\sqrt{-g} ~\left(c_1
  \, R\Box R  +c_2\,  R_{pq} \Box R^{pq}  +c_3 \, R^3   + c_4
\,  RR_{pq} R^{pq}  +c_5 \, R_{pq} R^{p}_{\phantom{p} r}R^{qr} \right. \nonumber \\
& &  \qquad \left.
+c_6
\, R_{pq}R_{rs}R^{prqs}  + c_7 \, RR_{pqrs} R^{pqrs}
  + c_8 \, R_{pq}R^p_{\phantom{p} rst} R^{qrst }   + c_9 \, R_{pqrs}R^{pquv}
R^{rs}_{\phantom{rs} uv} + c_{10} \, R_{prqs} R^{p \phantom{u}
q}_{\phantom{p} u \phantom{q} v} R^{r u s v} \right).
\end{eqnarray}
\end{widetext}
Here, the coefficients $c_i$ depend on the field and are given in Table~\ref{tab:table1}. In Eq.~(\ref{EffAct_2}) the integrand is constituted by ten Riemann polynomials of order six (in the derivatives of the metric tensor) and rank zero (number of free indices). It should be noted that terms in $1/m^4$, $1/m^6$ ... are also present in the full expression of the renormalized effective action $W_\mathrm{ren}$ (see, e.g., Refs.~\cite{DeWitt:2003pm,Avramidi:1986mj,Avramidi:2000bm}) but here, because we assume a large enough mass $m$, we can neglect them. It is also important to recall that, in the large mass limit, the nonlocal contribution to $W_\mathrm{ren}$ associated with the quantum state of the massive field is not taken into account. So, the renormalized effective action $W_\mathrm{ren}$ is a purely geometrical object.

\begin{table*}[htbp!]
\centering
\caption{\label{tab:table2} The coefficients $d_i$ defining the expansion of the approximate RSET (\ref{ren_SET_gen}) on the FKWC basis.}
\begin{ruledtabular}
\begin{tabular}{ll|cccc}
\multicolumn{2}{c}{FKWC basis} & Coefficients $d_i$ & \quad Scalar field \quad & \quad Dirac field \quad & \quad  Proca field \quad \\
\multicolumn{2}{c} {} &   & \quad $(s=0)$ \quad & \quad $(s=1/2)$ \quad & \quad  $(s=1)$ \quad \\
\hline\hline
\multirow{2}{*}{$\mathcal{R}^2_{6,1}$}
& $(\Box R)_{;\mu\nu}$ & $d_1$ & $ \xi^2-(2/5)\xi+3/70 $  & 1/70     &  9/70  \\
& $\Box \Box R_{\mu \nu}$ & $d_2$ & $ -1/140 $  &  -1/28 &   -9/28 \\
\hline
\multirow{13}{*}{$\mathcal{R}^2_{\{2,0\}}$}
& $R R_{;\mu \nu}$ & $d_3$ & $  -6\,(\xi-1/6)[\xi^2-(1/3)\xi+1/30] $  & -1/120  &    -1/10 \\
& $(\Box R) R_{\mu \nu}$ & $d_4$ & $ -(\xi-1/6)(\xi-1/5)\, $  &  1/120 &   -7/30\\
& $R_{;p (\mu} R^{p}_{\phantom{p}\nu)}$ & $d_5$ & $ (1/15)(\xi-1/7) $  & 23/840  &   13/35 \\
& $R \Box R_{\mu \nu}$ & $d_6$ & $ (1/10)(\xi-1/6)  $  & 1/40  &    -7/60 \\
& $R_{p (\mu} \Box R^p_{\phantom{p} \nu)}$ & $d_7$ & $ 1/42 $  &  29/420 &   337/210 \\
& $R^{pq} R_{pq;(\mu \nu)}$ & $d_8$ & $ (1/15)(\xi-2/7) $  & -19/420  &   22/105 \\
& $R^{pq} R_{p (\mu ; \nu)q}$ & $d_9$ & $ 2/105 $  &  61/420  &    34/105 \\
& $R^{pq} R_{\mu\nu; pq}$ & $d_{10}$ & $ -1/70 $  & -11/105  &   -107/210 \\
& $R^{;pq}R_{p \mu q \nu}$ & $d_{11}$ & $ (2/15)(\xi-3/14) $  &    -1/105 &   1/21 \\
& $(\Box R^{pq})R_{p \mu q \nu}$ & $d_{12}$ & $ -1/105 $  & -17/210  &  -22/35 \\
& $R^{pq;r}_{\phantom{pq;r} (\mu} R_{|rqp|\nu)}$ & $d_{13}$ & $ 4/105 $  & 13/105  &  46/35 \\
& $R^{p \phantom{(\mu}; qr}_{\phantom{p } (\mu} R_{|pqr|\nu)}$ & $d_{14}$ & $ 2/35 $  &   16/105 &  116/105 \\
& $R^{pqrs} R_{pqrs ; (\mu \nu)}$ & $d_{15}$ & $ -(1/15)(\xi-3/14) $  & 1/210  &  -1/42 \\
\hline
\multirow{11}{*}{$\mathcal{R}^2_{\{1,1\}}$}
& $R_{;\mu} R_{;\nu}$ & $d_{16}$ & $ -6(\xi-1/4)(\xi-1/6)^2 $  & 0  &   -1/24 \\
& $R_{;p } R^p_{\phantom{p} (\mu;\nu)}$ & $d_{17}$ & $  -(1/5)(\xi-3/14) $  & 19/840  &   83/210 \\
& $R_{;p }R_{\mu\nu}^{\phantom{\mu\nu}; p}$ & $d_{18}$ & $ (1/5)(\xi-17/84)  $  & -1/420  &  -41/84 \\
& $R^{pq}_{\phantom{pq};\mu}R_{pq;\nu}$ & $d_{19}$ & $ (1/15)(\xi-1/4) $  &  0 &   31/60 \\
& $R^{pq}_{\phantom{pq};(\mu}R_{\nu)p;q}$ & $d_{20}$ & $ 0 $  & -1/60  & -14/15  \\
& $R^p_{\phantom{p} \mu;q} R_{p \nu}^{\phantom{p \nu};q}$ & $d_{21}$ & $ -1/210 $  & -1/140  &  221/210 \\
& $R^p_{\phantom{p} \mu;q} R^q_{\phantom{q} \nu;p}$ & $d_{22}$ & $ 1/42 $  & 3/35  &  113/210 \\
& $R^{pq;r} R_{rqp (\mu;\nu) }$ & $d_{23}$ & $ -1/105 $  & -1/21  &  5/21\\
& $R^{pq;r} R_{p \mu q \nu;r}$ & $d_{24}$ & $ -1/70 $  & -11/105 &   -107/210 \\
& $R^{pqrs}_{\phantom{pqrs};\mu} R_{pqrs ; \nu }$ & $d_{25}$ & $ -(1/15)(\xi-13/56) $  & 1/420  &   -17/840 \\
& $R^{pqr}_{\phantom{pqr}\mu;s}R_{pqr \nu}^{\phantom{pqr\nu};s}$ & $d_{26}$ & $ -1/70 $  & -4/105  &   -29/105 \\
\hline
\multirow{16}{*}{$\mathcal{R}^2_{6,3}$}
& $R^2 R_{\mu\nu}$ & $d_{27}$ & $ 3(\xi-1/6)^3 $  & -1/288  &  5/24 \\
& $R R_{p \mu} R^p_{\phantom{p}\nu}$ & $d_{28}$ & $ -(2/15)(\xi-1/6) $  & -7/360  &  -2/5 \\
& $R^{pq}R_{pq}R_{\mu\nu}$ & $d_{29}$ & $ -(1/30)(\xi-1/6) $  & 1/180  &   -31/60 \\
& $R^{pq}R_{p \mu}R_{q \nu}$ & $d_{30}$ & $ -2/315 $  & -1/252 &  1/21 \\
& $R R^{pq}R_{p \mu q \nu}$ & $d_{31}$ & $ (1/15)(\xi-1/6) $  & 11/360  &  -19/30 \\
& $R^{pr}R^q_{\phantom{q} r}R_{p \mu q \nu}$ & $d_{32}$ & $ 1/315 $  & 13/1260  &   33/35 \\
& $R^{pq}R^r_{\phantom{r} (\mu}R_{ |rqp|  \nu) }$ & $d_{33}$ & $ 1/315 $  &  97/1260 &   -139/105 \\
& $R R^{pqr}_{\phantom{pqr}\mu}R_{pqr \nu}$ & $d_{34}$ & $ (1/15)(\xi-1/6) $  & 7/720  &   1/5 \\
& $R_{\mu\nu}R^{pqrs} R_{pqrs}$ & $d_{35}$ & $ (1/30)(\xi-1/6) $  & 7/1440  &  1/10 \\
& $R^p_{\phantom{p}  (\mu}R^{qrs}_{\phantom{qrs} |p|}R_{ |qrs| \nu) }$ & $d_{36}$ & $ -4/315 $  &  -73/1260  &  -74/105 \\
& $R^{pq}R^{rs}_{\phantom{rs} p\mu}R_{rs q \nu}$ & $d_{37}$ & $  -2/315 $  &  19/504 &  -5/42 \\
& $R_{pq}R^{p r q s}R_{r \mu s \nu}$ & $d_{38}$ & $ 4/315 $  &  73/1260 &  74/105 \\
& $R_{pq}R^{p rs}_{\phantom{p rs}\mu}R^q_{\phantom{q} rs \nu}$ & $d_{39}$ & $  -1/315 $  & -97/1260  &  -71/105 \\
& $R^{pq rs}R_{pq t \mu }R_{rs \phantom{t} \nu}^{\phantom{rs} t}$ & $d_{40}$ & $ 2/315 $  &  73/2520 &   37/105 \\
& $R^{p r q s}R^t_{\phantom{t} pq \mu}R_{t rs \nu}$ & $d_{41}$ & $ 4/63 $  &  239/1260 &   97/105 \\
& $R^{pqr}_{\phantom{pqr} s } R_{pqr t}R^{s \phantom{\mu} t}_{\phantom{s} \mu \phantom{t} \nu}$ & $d_{42}$ & $ -2/315 $  &  -73/2520  &   -37/105 \\
\hline
\multirow{1}{*}{$\mathcal{R}^0_{6,1}$}
& $\Box \Box R$ & $d_{43}$ & $ -\xi^2+ (2/5)\, \xi-11/280 $  & 1/280  &  9/280 \\
\hline
\multirow{4}{*}{$\mathcal{R}^0_{\{2,0\}}$}
& $R\Box R$ & $d_{44}$ & $ 6(\xi-1/6)[\xi^2- (1/3)\, \xi+1/40] $  & -1/240  &  19/120 \\
& $R_{;p q} R^{pq}$ & $d_{45}$ & $ -(1/30)(\xi-3/14)  $  & 1/420  &  -1/84 \\
& $R_{pq} \Box R^{pq}$ & $d_{46}$ & $  -(1/15)(\xi-5/28) $  & 1/105  &   -223/420 \\
& $R_{pq ; rs}R^{prqs}$ & $d_{47}$ & $ (4/15)(\xi-1/7) $  &  3/70 &  79/105 \\
\hline
\multirow{4}{*}{$\mathcal{R}^0_{\{1,1\}}$}
& $R_{;p}R^{;p}$ & $d_{48}$ & $ 6[\xi^3-(13/24)\xi^2+(17/180)\xi-53/10080] $  & -1/672  &  163/1680 \\
& $R_{pq;r} R^{pq;r}$ & $d_{49}$ & $ -(1/15)(\xi-13/56) $  & 3/280  &   -17/56 \\
& $R_{pq;r} R^{pr;q}$ & $d_{50}$ & $  -1/420 $  & -1/280 &  11/420 \\
& $R_{pqrs;t} R^{pqrs;t}$ & $d_{51}$ & $ (1/15)(\xi-19/112) $  & 1/168  &  51/560\\
\hline
\multirow{8}{*}{$\mathcal{R}^0_{6,3}$}
& $R^3$ & $d_{52}$ & $  - (1/2)(\xi-1/6)^3 $  & 1/1728  &  -5/144 \\
& $RR_{pq} R^{pq}$ & $d_{53}$ & $ (1/60)(\xi-1/6) $  & -1/360 &   31/120 \\
& $R_{pq} R^{p}_{\phantom{p} r}R^{qr}$ & $d_{54}$ & $ 1/1890 $  &  -1/945 &  1/630 \\
& $R_{pq}R_{rs}R^{prqs}$ & $d_{55}$ & $ -1/630 $  & 1/315  &  -53/105 \\
& $RR_{pqrs} R^{pqrs}$ & $d_{56}$ & $ -(1/60)(\xi-1/6)  $  &   -7/2880 &  -1/20 \\
& $R_{pq}R^p_{\phantom{p} rst} R^{qrst}$ & $d_{57}$ & $ (2/15)(\xi-1/6)  $  & 7/360  &  2/5 \\
& $R_{pqrs}R^{pquv} R^{rs}_{\phantom{rs} uv}$ & $d_{58}$ & $  -(1/15)(\xi-47/252) $  &  -61/15120  &  -263/2520\\
& $R_{prqs} R^{p \phantom{u} q}_{\phantom{p} u \phantom{q} v} R^{r u s v}$ & $d_{59}$ & $  -(4/15)(\xi-41/252) $  & -43/1512  & -106/315
\end{tabular}
\end{ruledtabular}
\end{table*}

By functional derivation of the effective action (\ref{EffAct_2}) with respect to the metric tensor $g_{\mu \nu}$,
we obtain a purely geometrical approximation for the RSET associated with the massive fields. We have
\begin{equation}\label{ST_EffAct 1}
 \langle  ~T^{\mu\nu
} ~ \rangle_{\mathrm{ren}} =\frac{2}{ \sqrt{-g}} \frac{\delta
W_{\mathrm{ren}}} {\delta
g_{\mu\nu }}
\end{equation}
which can be written explicitly \cite{Decanini:2007zz}
\begin{widetext}
\begin{eqnarray}\label{ren_SET_gen}  && (96 \pi^2 m^2)   \langle  ~T_{\mu\nu
} ~  \rangle_{\mathrm{ren}}    = d_1 \, (\Box R)_{;\mu\nu}
+d_2 \,  \Box \Box R_{\mu \nu} + d_3 \,  R R_{;\mu \nu}+ d_4 \, (\Box R) R_{\mu \nu} + d_5
\,R_{;p (\mu} R^{p}_{\phantom{p}\nu)}  + d_6 \, R \Box R_{\mu \nu} \nonumber \\
&& \qquad \quad + d_7 \, R_{p (\mu}
\Box
R^p_{\phantom{p} \nu)}  + d_8 \,R^{pq} R_{pq;(\mu \nu)}
 + d_9 \, R^{pq} R_{p (\mu ; \nu)q} + d_{10} \, R^{pq} R_{\mu\nu; pq}
  + d_{11} \,R^{;pq}R_{p \mu q \nu}  + d_{12} \, (\Box R^{pq})R_{p \mu q \nu} \nonumber \\
&& \qquad \quad +
 d_{13} \,R^{pq;r}_{\phantom{pq;r} (\mu} R_{|rqp|
\nu)}  + d_{14}  \,R^{p \phantom{(\mu}; qr}_{\phantom{p } (\mu} R_{|pqr|
\nu)} + d_{15} \,R^{pqrs} R_{pqrs ; (\mu \nu)
}  + d_{16} \, R_{;\mu} R_{;\nu} + d_{17} \,
R_{;p } R^p_{\phantom{p} (\mu;\nu)} \nonumber \\
&& \qquad \quad + d_{18} \,R_{;p }R_{\mu\nu}^{\phantom{\mu\nu}; p}
+ d_{19} \, R^{pq}_{\phantom{pq};\mu}R_{pq;\nu} + d_{20} \, R^{pq}_{\phantom{pq};(\mu}R_{\nu)p;q} + d_{21} \,
R^p_{\phantom{p} \mu;q} R_{p \nu}^{\phantom{p \nu};q} +
 d_{22} \,R^p_{\phantom{p} \mu;q} R^q_{\phantom{q} \nu;p} + d_{23}\,
R^{pq;r} R_{rqp (\mu;\nu) } \nonumber \\
&& \qquad \quad + d_{24} \, R^{pq;r} R_{p \mu q \nu;r}
+ d_{25} \,R^{pqrs}_{\phantom{pqrs};\mu} R_{pqrs ; \nu }+ d_{26} \, R^{pqr}_{\phantom{pqr}\mu;s}R_{pqr \nu}^{\phantom{pqr
\nu};s}   + d_{27} \, R^2 R_{\mu\nu} + d_{28} \, R R_{p \mu} R^p_{\phantom{p}\nu}
+ d_{29} \, R^{pq}R_{pq}R_{\mu\nu} \nonumber \\
&& \qquad \quad  + d_{30} \, R^{pq}R_{p \mu}R_{q \nu}
+ d_{31} \, R R^{pq}R_{p \mu q \nu}     + d_{32} \,R^{pr}R^q_{\phantom{q} r}R_{p \mu q \nu} + d_{33} \, R^{pq}R^r_{\phantom{r} (\mu}R_{ |rqp|  \nu) }
+ d_{34} \, R R^{pqr}_{\phantom{pqr}\mu }R_{pqr \nu} \nonumber \\
&& \qquad \quad
 + d_{35} \,R_{\mu\nu}R^{pqrs} R_{pqrs }
 + d_{36}\, R^p_{\phantom{p}  (\mu}R^{qrs}_{\phantom{qrs} |p|
}R_{ |qrs| \nu) } + d_{37} \, R^{pq}R^{rs}_{\phantom{rs} p\mu}R_{rs q \nu}+ d_{38} \, R_{pq}R^{p r q
s}R_{r \mu s \nu} + d_{39} \, R_{pq}R^{p rs}_{\phantom{p rs}\mu}R^q_{\phantom{q} rs \nu} \nonumber \\
&& \qquad \quad + d_{40} \,R^{pq rs}R_{pq t \mu }R_{rs \phantom{t}
\nu}^{\phantom{rs} t}
 + d_{41} \, R^{p r q s}R^t_{\phantom{t}
pq \mu}R_{t rs \nu}  + d_{42} \,  R^{pqr}_{\phantom{pqr} s } R_{pqr t}R^{s
\phantom{\mu} t}_{\phantom{s} \mu \phantom{t} \nu} \nonumber \\
&&  \qquad
+ g_{\mu\nu} [  d_{43} \, \Box \Box R     + d_{44} \, R\Box R
+ d_{45} \, R_{;p q}
R^{pq} + d_{46} \,  R_{pq} \Box R^{pq} + d_{47} \, R_{pq ; rs}R^{prqs} + d_{48} \,R_{;p}R^{;p}+ d_{49} \, R_{pq;r} R^{pq;r} \nonumber \\
&& \qquad  \quad + d_{50} \, R_{pq;r}
R^{pr;q} + d_{51} \,R_{pqrs;t} R^{pqrs;t}
     + d_{52} \, R^3   + d_{53} \, RR_{pq} R^{pq} + d_{54} \,R_{pq}
R^{p}_{\phantom{p} r}R^{qr}  + d_{55} \, R_{pq}R_{rs}R^{prqs}  \nonumber \\
&& \qquad  \quad + d_{56} \,
RR_{pqrs} R^{pqrs}
 + d_{57} \,R_{pq}R^p_{\phantom{p} rst} R^{qrst } + d_{58} \, R_{pqrs}R^{pquv}
R^{rs}_{\phantom{rs} uv}  + d_{59} \, R_{prqs} R^{p \phantom{u}
q}_{\phantom{p} u \phantom{q} v} R^{r u s v} ].
\end{eqnarray}
\end{widetext}
This formula is the basic cornerstone of our calculations and it is necessary to comment on it briefly. It should be noted that the expression of the RSET has been written in an ``irreducible form". Indeed, if we do not carefully take into account the symmetries of the Riemann tensor as well as the Bianchi identities, the metric variation of the renormalized effective action (\ref{EffAct_2}) could lead to an expression with many of the terms which are linearly dependent in a nontrivial way. As a consequence, the resulting expression contains too many terms (this is the case in most of the articles dealing with the DeWitt-Schwinger approximation) and, in practice, this increases significantly the size of calculations. For that reason, in Ref.~\cite{Decanini:2007zz}, we have expanded the RSET on a standard basis constituted by Riemann polynomials of order six in the derivatives of the metric tensor and constructed from group theoretical considerations by Fulling, King, Wybourne and
Cummings (FKWC) \cite{Fulling:1992vm}. This basis is described in Secs.~2.1 and 2.2 of Ref.~\cite{Decanini:2007zz} and displayed in Table~\ref{tab:table2} where we use, furthermore, the FKWC notations ${\cal R}^r_{s,q}$ and ${\cal
R}^r_{\lbrace{\lambda_1 \dots \rbrace}}$ to denote the various subspaces of the space of Riemann polynomials of rank r (see Ref.~\cite{Fulling:1992vm} for more precision). It should be finally noted that there also exist two geometrical identities [see Eqs.~(3.21) and (3.22) in Ref.~\cite{Decanini:2007zz}] coming from a topological and a geometrical constraint due to the four-dimensional nature of spacetime. We could have used them in order to eliminate two other terms in (\ref{ren_SET_gen}) but we have chosen to work with the FKWC basis of Riemann polynomials of order six which can be used in any dimension.

\section{Approximate renormalized stress tensors in Kerr-Newman spacetime}  \label{SectionII}

In this section, we consider the massive scalar field, the massive Dirac field and the Proca field in the Kerr-Newman spacetime and we provide, in the large mass limit, the associated explicit expressions of the approximate RSET $\langle T{}^\mu{}_\nu \rangle {}_\mathrm{ren}$. We work with Boyer-Lindquist coordinates and the spacetime metric then takes the form \cite{Misner:1974qy}
\begin{widetext}
\begin{equation} \label{KerrNewman_metric}
ds^2 = -\left(\frac{\Delta - a^2 \sin^2\theta}{\Sigma}\right) \, dt^2 -\frac{2 a \sin^2\theta (r^2 + a^2 - \Delta)}{\Sigma} \, dt d\varphi + \frac{(r^2+a^2)^2 - a^2 \Delta \sin^2\theta}{\Sigma} \sin^2\theta \, d\varphi^2 + \frac{\Sigma}{\Delta} \, dr^2 + \Sigma \, d\theta^2
\end{equation}
\end{widetext}
where $\Delta = r^2 - 2 M r + a^2  + Q^2$ and $\Sigma = r^2 + a^2 \cos^2\theta$. Here $M$, $Q$ and $J = a M$ are the mass, the charge and the angular momentum of the black hole while $a$ is the so-called rotation parameter and we assume $M^2 \ge a^2 + Q^2$. The outer horizon is located at $r_+ = M + \sqrt{M^2 - (a^2+Q^2)}$, the largest root of $\Delta$.

By using the general formula (\ref{ren_SET_gen}) and Table~\ref{tab:table2} in connection with (\ref{KerrNewman_metric}), we can obtain explicitly $\langle T{}^\mu{}_\nu \rangle {}_\mathrm{ren}$. Of course, due to the complexity of the Kerr-Newman metric, the calculations involved cannot be done by hand. For this reason, we have written the package \textit{SETArbitraryST} (available upon request from the first author). It is based on the suite of \textit{Mathematica} packages \textit{xAct} \cite{MartinGarcia} which permits us to perform tensor algebra very efficiently.

\begin{widetext}

The explicit expressions of the nonzero components of $\langle T{}^\mu{}_\nu \rangle {}_\mathrm{ren}$ can be written in the same form for the different massive fields. We have
\begin{eqnarray}\label{KN_SETtt}
&&\langle T{}^t{}_t \rangle {}_\mathrm{ren} =\; \frac{M^2 r^{10}}{40320 \pi ^2 m^2 \left(r^2 + a^2 \cos^2\theta \right)^9}
\sum_{p=0}^{5}
\left\lbrace
\sum_{q=0}^{3}
A^{tt}{}_{p,q} \left[ \theta , M/r \right]
\left(\frac{Q^2}{M^2}\right)^{q}
\right\rbrace \left(\frac{a}{r}\right)^{2p} \text{,}
\end{eqnarray}
\begin{eqnarray}\label{KN_SETrr}
&&\langle T{}^r{}_r \rangle {}_\mathrm{ren} =\; \frac{M^2 r^{10}}{40320 \pi ^2 m^2 \left(r^2 + a^2 \cos^2\theta \right)^9}
\sum_{p=0}^{5}
\left\lbrace
\sum_{q=0}^{3}
A^{r r}{}_{p,q} \left[ \theta , M/r \right]
\left(\frac{Q^2}{M^2}\right)^{q}
\right\rbrace \left(\frac{a}{r}\right)^{2p} \text{,}
\end{eqnarray}
\begin{eqnarray}\label{KN_SETthetatheta}
&&\langle T{}^\theta{}_\theta \rangle {}_\mathrm{ren} =\; \frac{M^2 r^{10}}{40320 \pi ^2 m^2 \left(r^2 + a^2 \cos^2\theta \right)^9}
\sum_{p=0}^{5}
\left\lbrace
\sum_{q=0}^{3}
A^{\theta \theta}{}_{p,q} \left[ \theta , M/r \right]
\left(\frac{Q^2}{M^2}\right)^{q}
\right\rbrace \left(\frac{a}{r}\right)^{2p} \text{,}
\end{eqnarray}
\begin{eqnarray}\label{KN_SETphiphi}
&&\langle T{}^\varphi{}_\varphi \rangle {}_\mathrm{ren} =\; \frac{M^2 r^{10}}{40320 \pi ^2 m^2 \left(r^2 + a^2 \cos^2\theta \right)^9}
\sum_{p=0}^{5}
\left\lbrace
\sum_{q=0}^{3}
A^{\varphi \varphi}{}_{p,q} \left[ \theta , M/r \right]
\left(\frac{Q^2}{M^2}\right)^{q}
\right\rbrace \left(\frac{a}{r}\right)^{2p} \text{,}
\end{eqnarray}
\begin{eqnarray}\label{KN_SETtphi}
&&\langle T{}^t{}_\varphi \rangle {}_\mathrm{ren} =\; \frac{M^2 r^{11} \sin^2\theta}{20160 \pi ^2 m^2 \left(r^2 + a^2 \cos^2\theta \right)^9}
\sum_{p=0}^{5}
\left\lbrace
\sum_{q=0}^{3}
A^{t \varphi}{}_{p,q} \left[ \theta , M/r \right]
\left(\frac{Q^2}{M^2}\right)^{q}
\right\rbrace \left(\frac{a}{r}\right)^{2p+1} \text{,}
\end{eqnarray}
\begin{eqnarray}\label{KN_SETphit}
&&\langle T{}^\varphi{}_t \rangle {}_\mathrm{ren} =\; \frac{M^2 r^{9}}{20160 \pi ^2 m^2 \left(r^2 + a^2 \cos^2\theta \right)^9}
\sum_{p=0}^{4}
\left\lbrace
\sum_{q=0}^{3}
A^{\varphi t}{}_{p,q} \left[ \theta , M/r \right]
\left(\frac{Q^2}{M^2}\right)^{q}
\right\rbrace \left(\frac{a}{r}\right)^{2p+1} \text{,}
\end{eqnarray}
\begin{eqnarray}\label{KN_SETrtheta}
&&\langle T{}^r{}_\theta \rangle {}_\mathrm{ren} =\; \frac{M^2 r^{11} \sin\theta \cos\theta}{360 \pi ^2 m^2 \left(r^2 + a^2 \cos^2\theta \right)^9}
\sum_{p=0}^{4}
\left\lbrace
\sum_{q=0}^{3}
A^{r \theta}{}_{p,q} \left[ \theta , M/r \right]
\left(\frac{Q^2}{M^2}\right)^{q}
\right\rbrace \left(\frac{a}{r}\right)^{2p+2} \text{,}
\end{eqnarray}
\begin{eqnarray}\label{KN_SETthetar}
&&\langle T{}^\theta{}_r \rangle {}_\mathrm{ren} =\; \frac{M^2 r^{9} \sin\theta \cos\theta}{360 \pi ^2 m^2 \left(r^2 + a^2 \cos^2\theta \right)^9}
\sum_{p=0}^{3}
\left\lbrace
\sum_{q=0}^{2}
A^{\theta r}{}_{p,q} \left[ \theta , M/r \right]
\left(\frac{Q^2}{M^2}\right)^{q}
\right\rbrace \left(\frac{a}{r}\right)^{2p+2} .
\end{eqnarray}
Here, the coefficients $A^{\mu \nu}{}_{p,q}$ are polynomials of the variables $\cos^2\theta$ and $M/r$ with coefficients depending on the field. The interested reader can find these coefficients for the massive scalar field, the massive Dirac field and the Proca field in the subsections \ref{KN_SETScalar_coeff}, \ref{KN_SETDirac_coeff} and \ref{KN_SETProca_coeff} of the Appendix.

\end{widetext}

\section{Special cases : approximate renormalized stress tensors in Schwarzschild, Reissner-Nordstr\"om and Kerr spacetimes} \label{SectionIII}

The structure of the expressions (\ref{KN_SETtt})--(\ref{KN_SETthetar}) permits us to recover very quickly the results corresponding to the special cases of the Schwarzschild, Reissner-Nordstr\"om and Kerr spacetimes.

By taking $Q=0$ and $a=0$ we recover the results obtained in Schwarzschild spacetime by Frolov and Zelnikov (see Ref.~\cite{Frolov:1982fr} and Sec.~11.3.7 of Ref.~\cite{FrolovNovikov:1998}). The nonzero components of $\langle T{}^\mu{}_\nu \rangle {}_\mathrm{ren}$ are
\begin{eqnarray}\label{S_SETtt}
&&\langle T{}^t{}_t \rangle {}_\mathrm{ren} =\; \frac{M^2}{40320 \pi ^2 m^2 r^8}
\, A^{tt}{}_{0,0} \left[ \theta , M/r \right] \text{,}
\end{eqnarray}
\begin{eqnarray}\label{S_SETrr}
&&\langle T{}^r{}_r \rangle {}_\mathrm{ren} =\; \frac{M^2}{40320 \pi ^2 m^2 r^8}
\, A^{r r}{}_{0,0} \left[ \theta , M/r \right] \text{,}
\end{eqnarray}
\begin{eqnarray}\label{S_SETthetatheta}
&&\langle T{}^\theta{}_\theta \rangle {}_\mathrm{ren} =\; \frac{M^2}{40320 \pi ^2 m^2 r^8}
\, A^{\theta \theta}{}_{0,0} \left[ \theta , M/r \right] \text{,}
\end{eqnarray}
\begin{eqnarray}\label{S_SETphiphi}
&&\langle T{}^\varphi{}_\varphi \rangle {}_\mathrm{ren} =\; \frac{M^2}{40320 \pi ^2 m^2 r^8}
\, A^{\varphi \varphi}{}_{0,0} \left[ \theta , M/r \right] \text{,}
\end{eqnarray}
where the coefficients $A^{\mu \nu}{}_{0,0}$ do not depend explicitly of $\theta$.
Due to the spherical symmetry of the Schwarzschild black hole, $A^{\theta \theta}{}_{0,0} = A^{\varphi \varphi}{}_{0,0}$ and, as a consequence, $\langle T{}^\theta{}_\theta \rangle {}_\mathrm{ren} = \langle T{}^\varphi{}_\varphi \rangle {}_\mathrm{ren}$. It should be noted that, for the Dirac field, Frolov and Zelnikov have forgotten a multiplicative factor $1/2$. The absence of this factor seems to be due to an error of these authors in their construction of the effective action for the massive Dirac field from the (bosonic) Lichnerowicz operator. This error does not exist in Refs.~\cite{Avramidi:1986mj,Avramidi:2000bm} and in Table~\ref{tab:table1}.

Similarly, by putting $a=0$ into the expressions (\ref{KN_SETtt})--(\ref{KN_SETthetar}), we recover the results obtained in Reissner-Nordstr\"om spacetime by Anderson, Hiscock and Samuel \cite{Anderson:1994hg} (for the scalar field) and by Matyjasek \cite{Matyjasek:1999an} (for the Dirac and Proca fields). The nonzero components of $\langle T{}^\mu{}_\nu \rangle {}_\mathrm{ren}$ are
\begin{eqnarray}\label{RN_SETtt}
&&\langle T{}^t{}_t \rangle {}_\mathrm{ren} =\; \frac{M^2}{40320 \pi ^2 m^2 r^8}
\sum_{q=0}^{3}
A^{tt}{}_{0,q} \left[ \theta , M/r \right]
\left(\frac{Q^2}{M^2}\right)^{q} \text{,} \nonumber \\
\end{eqnarray}
\begin{eqnarray}\label{RN_SETrr}
&&\langle T{}^r{}_r \rangle {}_\mathrm{ren} =\; \frac{M^2}{40320 \pi ^2 m^2 r^8}
\sum_{q=0}^{3}
A^{r r}{}_{0,q} \left[ \theta , M/r \right]
\left(\frac{Q^2}{M^2}\right)^{q} \text{,} \nonumber \\
\end{eqnarray}
\begin{eqnarray}\label{RN_SETthetatheta}
&&\langle T{}^\theta{}_\theta \rangle {}_\mathrm{ren} =\; \frac{M^2}{40320 \pi ^2 m^2 r^8}
\sum_{q=0}^{3}
A^{\theta \theta}{}_{0,q} \left[ \theta , M/r \right]
\left(\frac{Q^2}{M^2}\right)^{q} \text{,} \nonumber \\
\end{eqnarray}
\begin{eqnarray}\label{RN_SETphiphi}
&&\langle T{}^\varphi{}_\varphi \rangle {}_\mathrm{ren} =\; \frac{M^2}{40320 \pi ^2 m^2 r^8}
\sum_{q=0}^{3}
A^{\varphi \varphi}{}_{0,q} \left[ \theta , M/r \right]
\left(\frac{Q^2}{M^2}\right)^{q} \text{,} \nonumber \\
\end{eqnarray}
where the coefficients $A^{\mu \nu}{}_{0,q}$ do not depend explicitly of $\theta$. Due to the spherical symmetry of the Reissner-Nordstr\"om black hole, $A^{\theta \theta}{}_{0,q} = A^{\varphi \varphi}{}_{0,q}$ and, as a consequence, $\langle T{}^\theta{}_\theta \rangle {}_\mathrm{ren} = \langle T{}^\varphi{}_\varphi \rangle {}_\mathrm{ren}$.

\begin{widetext}

Finally, by putting $Q=0$ into the expressions (\ref{KN_SETtt})--(\ref{KN_SETthetar}), we can find the results in Kerr spacetime. The nonzero components of $\langle T{}^\mu{}_\nu \rangle {}_\mathrm{ren}$ are
\begin{eqnarray}\label{K_SETtt}
&&\langle T{}^t{}_t \rangle {}_\mathrm{ren} =\; \frac{M^2 r^{10}}{40320 \pi ^2 m^2 \left(r^2 + a^2 \cos^2\theta \right)^9}
\sum_{p=0}^{5}
A^{tt}{}_{p,0} \left[ \theta , M/r \right]
\left(\frac{a}{r}\right)^{2p} \text{,}
\end{eqnarray}
\begin{eqnarray}\label{K_SETrr}
&&\langle T{}^r{}_r \rangle {}_\mathrm{ren} =\; \frac{M^2 r^{10}}{40320 \pi ^2 m^2 \left(r^2 + a^2 \cos^2\theta \right)^9}
\sum_{p=0}^{5}
A^{r r}{}_{p,0} \left[ \theta , M/r \right]
\left(\frac{a}{r}\right)^{2p} \text{,}
\end{eqnarray}
\begin{eqnarray}\label{K_SETthetatheta}
&&\langle T{}^\theta{}_\theta \rangle {}_\mathrm{ren} =\; \frac{M^2 r^{10}}{40320 \pi ^2 m^2 \left(r^2 + a^2 \cos^2\theta \right)^9}
\sum_{p=0}^{5}
A^{\theta \theta}{}_{p,0} \left[ \theta , M/r \right]
\left(\frac{a}{r}\right)^{2p} \text{,}
\end{eqnarray}
\begin{eqnarray}\label{K_SETphiphi}
&&\langle T{}^\varphi{}_\varphi \rangle {}_\mathrm{ren} =\; \frac{M^2 r^{10}}{40320 \pi ^2 m^2 \left(r^2 + a^2 \cos^2\theta \right)^9}
\sum_{p=0}^{5}
A^{\varphi \varphi}{}_{p,0} \left[ \theta , M/r \right]
\left(\frac{a}{r}\right)^{2p} \text{,}
\end{eqnarray}
\begin{eqnarray}\label{K_SETtphi}
&&\langle T{}^t{}_\varphi \rangle {}_\mathrm{ren} =\; \frac{M^2 r^{11} \sin^2\theta}{20160 \pi ^2 m^2 \left(r^2 + a^2 \cos^2\theta \right)^9}
\sum_{p=0}^{5}
A^{t \varphi}{}_{p,0} \left[ \theta , M/r \right]
\left(\frac{a}{r}\right)^{2p+1} \text{,}
\end{eqnarray}
\begin{eqnarray}\label{K_SETphit}
&&\langle T{}^\varphi{}_t \rangle {}_\mathrm{ren} =\; \frac{M^2 r^{9}}{20160 \pi ^2 m^2 \left(r^2 + a^2 \cos^2\theta \right)^9}
\sum_{p=0}^{4}
A^{\varphi t}{}_{p,0} \left[ \theta , M/r \right]
\left(\frac{a}{r}\right)^{2p+1} \text{,}
\end{eqnarray}
\begin{eqnarray}\label{K_SETrtheta}
&&\langle T{}^r{}_\theta \rangle {}_\mathrm{ren} =\; \frac{M^2 r^{11} \sin\theta \cos\theta}{360 \pi ^2 m^2 \left(r^2 + a^2 \cos^2\theta \right)^9}
\sum_{p=0}^{4}
A^{r \theta}{}_{p,0} \left[ \theta , M/r \right]
\left(\frac{a}{r}\right)^{2p+2} \text{,}
\end{eqnarray}
\begin{eqnarray}\label{K_SETthetar}
&&\langle T{}^\theta{}_r \rangle {}_\mathrm{ren} =\; \frac{M^2 r^{9} \sin\theta \cos\theta}{360 \pi ^2 m^2 \left(r^2 + a^2 \cos^2\theta \right)^9}
\sum_{p=0}^{3}
A^{\theta r}{}_{p,0} \left[ \theta , M/r \right]
\left(\frac{a}{r}\right)^{2p+2} .
\end{eqnarray}

\end{widetext}

It should be noted that the approximate RSETs for the massive scalar field, the massive Dirac field and the Proca field in Kerr spacetime have been obtained a long time ago by Frolov and Zelnikov \cite{Frolov:1983ig,Frolov:1984ra}. At first sight, their results and ours are different because theirs are given in terms of the complex spin coefficient $\rho=-(r- i a \cos\theta)^{-1}$ and its powers. In our opinion, our formulas are clearer. Moreover, we have checked that both results are in agreement for the scalar and vector fields while, for the Dirac field, they differ by the multiplicative factor $1/2$ previously discussed.

\section{Conclusion}  \label{SectionIV}

In this article, we have obtained an analytical approximation for the RSET of the massive scalar, spinor and vector fields in Kerr-Newman spacetime. To our knowledge, no other result concerning this tensor in Kerr-Newman spacetime can be found within the literature. The \textit{Mathematica} package \textit{SETArbitraryST} which permits us to derive the expressions of the RSET $\langle T_{\mu \nu} \rangle {}_\mathrm{ren}$ in Kerr-Newman spacetime as well as another package which contains these explicit expressions are available upon request from the first author.

The approximate expressions obtained are based on the DeWitt-Schwinger expansion of the effective action associated with a massive quantum field. As a consequence, they do not take into account the quantum state of the field and are only valid in the large mass limit. In particular, it is important to note that they neglect the existence of supperradiance instabilities for massive fields in rotating black holes (see Refs.~\cite{Damour:1976kh,Zouros:1979iw,Detweiler:1980uk} for pioneering work on this subject and, e.g., Ref.~\cite{Furuhashi:2004jk} for a more recent article concerning more particularly the Kerr-Newman black hole) which seems quite reasonable because the instability time scale is very long in the large mass limit. We have also shown that our results permit us to recover those already existing in the literature for the RSET in Schwarzschild, Reissner-Nordstr\"om and Kerr spacetimes and we hope they will be useful to people who will want, in the future, to make the exact calculations by taking into account, in particular, the quantum state of the massive field. But, in our opinion, the complexity of our results leads us to think that an exact expression for the RSET of a massive field in Kerr-Newman spacetime is completely out of reach.

However, this should not prevent us from discussing the following fundamental question : what exact result(s) may be associated with our approximation? This is far from obvious and, here, we shall only provide a partial answer to this important question. Let us first consider the case of Schwarzschild spacetime. It is well known that, in this gravitational background, three Hadamard vacua are physically relevant (see Ref.~\cite{Candelas:1980zt} and references therein): the so-called Hartle-Hawking ($| H \rangle$), Unruh ($| U \rangle$) and Boulware ($| B \rangle$) vacua. In Ref~\cite{Anderson:1994hg}, it has been shown that the DeWitt-Schwinger approximation (\ref{S_SETtt})--(\ref{S_SETphiphi}) is a good approximation  of $\langle H| T^\mu{}_\nu |H \rangle {}_\mathrm{ren}$ (i.e., of the exact RSET in the Hartle-Hawking vacuum) for small and intermediate values of $s=r/(2M)-1$. As noted by Frolov and Zelnikov \cite{Frolov:1984ra}, the differences between this mean value and the mean values in the Unruh vacuum $| U \rangle$ and the Boulware vacuum $| B \rangle$ are proportional to the factor $\exp( - m/T_\mathrm{BH})$ and this difference can be neglected everywhere except close to the horizon. As a consequence, the DeWitt-Schwinger approximation (\ref{S_SETtt})--(\ref{S_SETphiphi}) is certainly a good approximation of $\langle U| T^\mu{}_\nu |U \rangle {}_\mathrm{ren}$ and $\langle B| T^\mu{}_\nu |B \rangle {}_\mathrm{ren}$ for intermediate values of $s$. Similar considerations apply in the Reissner-Nordstr\"om spacetime. By contrast, in Kerr and Kerr-Newman spacetimes the problem is much more complicated due, in particular, to the superradiant modes (see, e.g., Ref.~\cite{Ottewill:2000qh}) and to the nonexistence of Hadamard states which respect the symmetries of the spacetime and are regular everywhere \cite{Kay:1988mu}. However, it seems that these difficulties are naturally eliminated for the fermionic fields \cite{Casals:2012es} and can be, in some sense, circumvented for bosonic fields \cite{Frolov:1989jh,Ottewill:2000qh,Ottewill:2000yr,Duffy:2005mz} and, in fact, one can consider that nonconventional Hartle-Hawking, Unruh and Boulware vacua exist in Kerr spacetime (and probably in Kerr-Newman spacetime). In our opinion, the DeWitt-Schwinger approximation (\ref{KN_SETtt})--(\ref{KN_SETthetar}) is certainly a good approximation of the RSETs associated with these nonconventional vacua, but, of course, the region of space where this approximation can be used necessarily depends
on the considered vacuum.

\begin{table}[t]
\caption{\label{tab:table3} The coefficients $\alpha_i$ and $\beta_i$ appearing in the shift formulas (\ref{ShiftM_exp}) and (\ref{ShiftJ_exp}). }
\begin{ruledtabular}
\begin{tabular}{cccc}
      & \quad Scalar field  \quad & \quad Dirac field  \quad &   Proca field  \\
           & \quad $(s=0)$  \quad & \quad $(s=1/2)$  \quad &   $(s=1)$  \\
\hline
$\alpha_0$ & $ -392(9\xi-2)$  & 196 & -1176   \\
$\alpha_1$ & $ 14(63\xi-16) $  & 7 & 210 \\
$\alpha_2$ & $ -66$  & -162 & 978 \\
\hline
$\beta_0$ & $-60(84\xi-17) $  & 480 & -1980 \\
$\beta_1$ & $-9(252\xi-53) $  & 180 & -837 \\
$\beta_2$ & $ 20(714\xi-145) $  & -640 & 19580
\end{tabular}
\end{ruledtabular}
\end{table}

In the absence of exact results in Kerr-Newman spacetime, our approximate RSETs could be very helpful, in particular to study the backreaction of massive quantum fields on this gravitational background. In Refs.~\cite{Hiscock:1997jt,Taylor:1999ic,Anderson:2000pg,Matyjasek:2006nu,Piedra:2009pf,Piedra:2010ur} which deal with massive field theories on the Schwarzschild and Reissner-Nordstr\"om black holes, some aspects of the backreaction problem have been considered. Unfortunately, here, for the Kerr-Newman black hole, due to the complexity of the RSET, it seems impossible to find self-consistent solutions to the semiclassical Einstein equations (\ref{SCEinsteinEq}). However, it is possible to simplify considerably the backreaction problem by limiting us to the determination of the shift in mass and angular momentum of the black hole (measured by a distant observer) due to the RSET. Such an approach has already been considered by Frolov and Zelnikov for the Kerr black hole \cite{Frolov:1983ig,Frolov:1984ra} and its extension to the Kerr-Newman black hole is tractable and permits us to emphasize the role of the black hole charge. For a stationary axisymmetric black hole, we recall that the mass $M_D$ and the angular momentum $J_D$ of the black hole dressed with a quantum field can be expressed in terms of its mass $M$ and its angular momentum $J$ by (see, e.g., Ref.~\cite{Bardeen:1973gs})
\begin{equation}\label{ShiftM}
M_D-M= 2\int_{\cal S} \left(\langle T{}^\mu{}_\nu \rangle {}_\mathrm{ren} -\frac{1}{2} g{}^\mu{}_\nu \langle T{}^\rho{}_\rho \rangle {}_\mathrm{ren}  \right) \xi^\nu d{\cal S}_\mu
\end{equation}
and
\begin{equation}\label{ShiftJ}
J_D-J=-\int_{\cal S}  \langle T{}^\mu{}_\nu \rangle {}_\mathrm{ren}   \psi^\nu d{\cal S}_\mu
\end{equation}
where $\langle T_{\mu \nu} \rangle {}_\mathrm{ren}$ is the RSET of the quantum field, $\xi^\mu=(\partial / \partial t)^\mu$ and $\psi^\mu = (\partial / \partial \varphi)^\mu$ denote the two Killing vectors of the Kerr-Newman black hole, ${\cal S}$ is any spacelike hypersurface that extends from the outer horizon at $r_+$ to spatial infinity and $d{\cal S}_\mu$ is the associated surface element. In the following, we take for ${\cal S}$ the hypersurface defined by $t= \mathrm{const}$ and we have therefore $d{\cal S}_\mu= - (r^2+a^2 \cos^2 \theta) \sin \theta  dr d\theta d\varphi \, (dt)_\mu$. By using the expressions (\ref{KN_SETtt}) - (\ref{KN_SETthetar}) and assuming furthermore $a \ll M$ and $Q \ll M$ (for arbitrary values of the parameters $a$ and $Q$, the expressions obtained are too complicated to be interesting), we obtain
\begin{eqnarray}\label{ShiftM_exp}
& & M_D-M= \frac{M}{336 \times 7! \pi m^2 M^4} \nonumber \\
&& \times \left\lbrace \alpha_0+ \alpha_1 \left( \frac{a}{M} \right)^2  + \alpha_2 \left( \frac{Q}{M} \right)^2 + \dots  \right\rbrace
\end{eqnarray}
and
\begin{eqnarray}\label{ShiftJ_exp}
& & J_D-J= \frac{1}{480 \times 7! \pi m^2 M^2}  \left( \frac{a}{M} \right)\nonumber \\
&& \times  \left\lbrace \beta_0+ \beta_1 \left( \frac{a}{M} \right)^2  + \beta_2 \left( \frac{Q}{M} \right)^2  + \dots \right\rbrace
\end{eqnarray}
where the dots denote terms of fourth-order [i.e., in $(a/M)^4$ or in $(Q/M)^4$ or in $a^2Q^2/M^4$]. The coefficients $\alpha_i$ and $\beta_i$ appearing in Eqs.~(\ref{ShiftM_exp}) and (\ref{ShiftJ_exp}) are given in Table~\ref{tab:table3}. For $Q=0$ (i.e., for the Kerr black hole), our results correct some errors made in Refs.~\cite{Frolov:1983ig,Frolov:1984ra} for the Dirac field (all the coefficients $\alpha_i$ and $\beta_i$ are incorrect) but also for the scalar and vector fields (the coefficient $\alpha_1$ is incorrect).

Our results could be also used to study the quasinormal modes of the Kerr-Newman black hole dressed by a massive quantum field. Similar problems have been considered in Refs.~\cite{Piedra:2009pf,Piedra:2010ur} for spherically symmetric black holes. The extension to the Kerr-Newman black hole is far from obvious, not only because of the RSET complexity but also because the subject of massive quasinormal modes in this spacetime is rather difficult and has been little studied (see, however, for recent articles dealing with this subject and which could be helpful, Ref.~\cite{Berti:2005eb} where the uncharged massless scalar field is considered and Ref.~\cite{Konoplya:2013rxa} where the charged massive scalar field is studied).

\begin{acknowledgments}

We wish to thank Yves D\'ecanini, Mohamed Ould El Hadj and Julien Queva for various discussions and the ``Collectivit\'e Territoriale de Corse" for its support through the COMPA project.

\end{acknowledgments}

\appendix

\begin{widetext}

\section{Coefficients $A^{\mu\nu}{}_{p,q} \left[ \theta , M/r \right]$} \label{KN_SET_coeff}

\subsection{Massive scalar field} \label{KN_SETScalar_coeff}

For the massive scalar field, the coefficients $A^{t t}{}_{p,q}$ appearing in the expression (\ref{KN_SETtt}) of $\langle T{}^t{}_t \rangle {}_\mathrm{ren}$ are

\begin{subequations}\label{KN_SETttScalar_coeff}
\begin{eqnarray}
&&
A^{t t}{}_{0,0} \left[ \theta , M/r \right] =   180 (112 \xi -25)  - 8 (5544 \xi -1237)  \left( M/r\right)
\\&&
A^{t t}{}_{0,1} \left[ \theta , M/r \right] =   1080  - 192 (294 \xi -41)  \left( M/r\right)+ 4 (36456 \xi -6845)  \left( M/r\right)^2
\\&&
A^{t t}{}_{0,2} \left[ \theta , M/r \right] =   8 (5096 \xi -883)  \left( M/r\right)^2- 48 (3164 \xi -613)  \left( M/r\right)^3
\\&&
A^{t t}{}_{0,3} \left[ \theta , M/r \right] =   52 (882 \xi -179)  \left( M/r\right)^4
\\&&
A^{t t}{}_{1,0} \left[ \theta , M/r \right] =   -4860 (112 \xi -25) \cos^2\theta   + 288 \left[2 (2303 \xi -513) \cos^2\theta +14 \xi -3\right]  \left( M/r\right)
\\&&
A^{t t}{}_{1,1} \left[ \theta , M/r \right] =   -1080 \left(7 \cos^2\theta -12\right)  + 144 \left[5 (1344 \xi -275) \cos^2\theta -224 \xi -219\right]  \left( M/r\right)
\nonumber \\&& \quad\quad - 16 \left[7 (26556 \xi -5903) \cos^2\theta +2 (462 \xi -769)\right]  \left( M/r\right)^2
\\&&
A^{t t}{}_{1,2} \left[ \theta , M/r \right] =   -16 \left[49 (472 \xi -95) \cos^2\theta -16 (140 \xi +43)\right]  \left( M/r\right)^2
\nonumber \\&& \quad\quad + 16 \left[28 (4553 \xi -1016) \cos^2\theta +1120 \xi -1559\right]  \left( M/r\right)^3
\\&&
A^{t t}{}_{1,3} \left[ \theta , M/r \right] =   -4 \left[(109942 \xi -24683) \cos^2\theta +4 (455 \xi -421)\right]  \left( M/r\right)^4
\\&&
A^{t t}{}_{2,0} \left[ \theta , M/r \right] =   7560 (112 \xi -25) \cos^4\theta   - 144 \cos^2\theta  \left[(24836 \xi -5525) \cos^2\theta +18 (14 \xi -3)\right]  \left( M/r\right)
\\&&
A^{t t}{}_{2,1} \left[ \theta , M/r \right] =   -2160 \left(4 \cos^4\theta +6 \cos^2\theta -15\right)
\nonumber \\&& \quad\quad  - 144 \left[8 (546 \xi -149) \cos^4\theta -3 (672 \xi -173) \cos^2\theta +335\right]  \left( M/r\right)
\nonumber \\&& \quad\quad + 24 \cos^2\theta  \left[5 (43680 \xi -9941) \cos^2\theta +4 (1470 \xi +61)\right]  \left( M/r\right)^2
\\&&
A^{t t}{}_{2,2} \left[ \theta , M/r \right] =   -96 \left[180 \cos^4\theta +2 (784 \xi -227) \cos^2\theta -253\right]  \left( M/r\right)^2
\nonumber \\&& \quad\quad - 16 \cos^2\theta  \left[7 (20612 \xi -4833) \cos^2\theta +2 (4424 \xi +377)\right]  \left( M/r\right)^3
\\&&
A^{t t}{}_{2,3} \left[ \theta , M/r \right] =   4 \cos^2\theta  \left[(72310 \xi -17861) \cos^2\theta +56 (175 \xi +17)\right]  \left( M/r\right)^4
\\&&
A^{t t}{}_{3,0} \left[ \theta , M/r \right] =   7560 (112 \xi -25) \cos^6\theta   + 96 \cos^4\theta  \left[2 (8631 \xi -1916) \cos^2\theta -15 (14 \xi -3)\right]  \left( M/r\right)
\\&&
A^{t t}{}_{3,1} \left[ \theta , M/r \right] =   2160 \left(4 \cos^6\theta -24 \cos^4\theta +15 \cos^2\theta +10\right)
\nonumber \\&& \quad\quad - 240 \cos^2\theta  \left[7 (768 \xi -169) \cos^4\theta -3 (224 \xi +47) \cos^2\theta +318\right]  \left( M/r\right)
\nonumber \\&& \quad\quad - 16 \cos^4\theta  \left[(90636 \xi -20021) \cos^2\theta +6 (630 \xi -101)\right]  \left( M/r\right)^2
\\&&
A^{t t}{}_{3,2} \left[ \theta , M/r \right] =   16 \cos^2\theta  \left[(23128 \xi -5195) \cos^4\theta -24 (392 \xi -1) \cos^2\theta +2700\right]  \left( M/r\right)^2
\nonumber \\&& \quad\quad + 16 \cos^4\theta  \left[14 (1414 \xi -291) \cos^2\theta +3472 \xi -939\right]  \left( M/r\right)^3
\\&&
A^{t t}{}_{3,3} \left[ \theta , M/r \right] =   -4 \cos^4\theta  \left[(2506 \xi -9) \cos^2\theta +4 (455 \xi -253)\right]  \left( M/r\right)^4
\\&&
A^{t t}{}_{4,0} \left[ \theta , M/r \right] =   -4860 (112 \xi -25) \cos^8\theta   - 72 \cos^6\theta  \left[(1456 \xi -321) \cos^2\theta -20 (14 \xi -3)\right]  \left( M/r\right)
\\&&
A^{t t}{}_{4,1} \left[ \theta , M/r \right] =   1080 \cos^2\theta  \left(7 \cos^6\theta -12 \cos^4\theta -30 \cos^2\theta +40\right)
\nonumber \\&& \quad\quad + 720 \cos^4\theta  \left[2 (252 \xi -65) \cos^4\theta -(224 \xi -107) \cos^2\theta -39\right]  \left( M/r\right)
\nonumber \\&& \quad\quad + 4 \cos^6\theta  \left[(9240 \xi -2069) \cos^2\theta -8 (42 \xi -17)\right]  \left( M/r\right)^2
\\&&
A^{t t}{}_{4,2} \left[ \theta , M/r \right] =   -8 \cos^4\theta  \left[13 (392 \xi -151) \cos^4\theta -8 (560 \xi -503) \cos^2\theta -2364\right]  \left( M/r\right)^2
\\&&
A^{t t}{}_{4,3} \left[ \theta , M/r \right] = 0
\\&&
A^{t t}{}_{5,0} \left[ \theta , M/r \right] =   180 (112 \xi -25) \cos^{10}\theta
\\&&
A^{t t}{}_{5,1} \left[ \theta , M/r \right] =   -1080 \cos^4\theta  \left(\cos^2\theta -2\right) \left(\cos^4\theta -10 \cos^2\theta +10\right)
\\&&
A^{t t}{}_{5,2} \left[ \theta , M/r \right] = 0
\\&&
A^{t t}{}_{5,3} \left[ \theta , M/r \right] = 0 .
\end{eqnarray}
\end{subequations}

For the massive scalar field, the coefficients $A^{r r}{}_{p,q}$ appearing in the expression (\ref{KN_SETrr}) of $\langle T{}^r{}_r \rangle {}_\mathrm{ren}$ are

\begin{subequations}\label{KN_SETrrScalar_coeff}
\begin{eqnarray}
&&
A^{r r}{}_{0,0} \left[ \theta , M/r \right] =   -252 (32 \xi -7)  + 56 (216 \xi -47)  \left( M/r\right)
\\&&
A^{r r}{}_{0,1} \left[ \theta , M/r \right] =   216  + 128 (147 \xi -40)  \left( M/r\right)- 4 (8232 \xi -2081)  \left( M/r\right)^2
\\&&
A^{r r}{}_{0,2} \left[ \theta , M/r \right] =   -8 (1456 \xi -383)  \left( M/r\right)^2+ 112 (252 \xi -65)  \left( M/r\right)^3
\\&&
A^{r r}{}_{0,3} \left[ \theta , M/r \right] =   -4 (1638 \xi -421)  \left( M/r\right)^4
\\&&
A^{r r}{}_{1,0} \left[ \theta , M/r \right] =   252 (32 \xi -7) \left(31 \cos^2\theta -4\right)  - 2016 (124 \xi -27) \cos^2\theta   \left( M/r\right)
\\&&
A^{r r}{}_{1,1} \left[ \theta , M/r \right] =   -216 \left(3 \cos^2\theta -8\right)  - 288 \left[7 (186 \xi -41) \cos^2\theta -2 (91 \xi -27)\right]  \left( M/r\right)
\nonumber \\&& \quad\quad + 48 (9744 \xi -2053) \cos^2\theta   \left( M/r\right)^2
\\&&
A^{r r}{}_{1,2} \left[ \theta , M/r \right] =   8 \left[7 (2288 \xi -503) \cos^2\theta -2800 \xi +867\right]  \left( M/r\right)^2- 16 (16268 \xi -3303) \cos^2\theta   \left( M/r\right)^3
\\&&
A^{r r}{}_{1,3} \left[ \theta , M/r \right] =   4 (11326 \xi -2197) \cos^2\theta   \left( M/r\right)^4
\\&&
A^{r r}{}_{2,0} \left[ \theta , M/r \right] =   -3528 (32 \xi -7) \cos^2\theta  \left(11 \cos^2\theta -8\right)  + 1008 (336 \xi -73) \cos^4\theta   \left( M/r\right)
\\&&
A^{r r}{}_{2,1} \left[ \theta , M/r \right] =   -432 \left(6 \cos^4\theta -6 \cos^2\theta -5\right)
\nonumber \\&& \quad\quad + 96 \cos^2\theta  \left[(14826 \xi -3187) \cos^2\theta -6 (2107 \xi -444)\right]  \left( M/r\right) - 8 (44520 \xi -9851) \cos^4\theta   \left( M/r\right)^2
\\&&
A^{r r}{}_{2,2} \left[ \theta , M/r \right] =   -8 \cos^2\theta  \left[(50064 \xi -10603) \cos^2\theta -50064 \xi +10279\right]  \left( M/r\right)^2
\nonumber \\&& \quad\quad + 16 (7196 \xi -1709) \cos^4\theta   \left( M/r\right)^3
\\&&
A^{r r}{}_{2,3} \left[ \theta , M/r \right] =   -4 (3346 \xi -951) \cos^4\theta   \left( M/r\right)^4
\\&&
A^{r r}{}_{3,0} \left[ \theta , M/r \right] =   3528 (32 \xi -7) \cos^4\theta  \left(17 \cos^2\theta -20\right)  + 672 (84 \xi -19) \cos^6\theta   \left( M/r\right)
\\&&
A^{r r}{}_{3,1} \left[ \theta , M/r \right] =   -432 \cos^2\theta  \left(4 \cos^4\theta +6 \cos^2\theta -15\right)
\nonumber \\&& \quad\quad  - 32 \cos^4\theta  \left[(59010 \xi -13061) \cos^2\theta -18 (4025 \xi -893)\right]  \left( M/r\right) - 16 (4704 \xi -1049) \cos^6\theta   \left( M/r\right)^2
\\&&
A^{r r}{}_{3,2} \left[ \theta , M/r \right] =   8 \cos^4\theta  \left[(55664 \xi -12563) \cos^2\theta -68880 \xi +15649\right]  \left( M/r\right)^2+ 16 (1036 \xi -235) \cos^6\theta   \left( M/r\right)^3
\phantom{M/r/r}
\\&&
A^{r r}{}_{3,3} \left[ \theta , M/r \right] =   -4 (182 \xi -81) \cos^6\theta   \left( M/r\right)^4
\\&&
A^{r r}{}_{4,0} \left[ \theta , M/r \right] =   -252 (32 \xi -7) \cos^6\theta  \left(85 \cos^2\theta -112\right)  - 72 (392 \xi -87) \cos^8\theta   \left( M/r\right)
\\&&
A^{r r}{}_{4,1} \left[ \theta , M/r \right] =   216 \cos^4\theta  \left(3 \cos^4\theta -28 \cos^2\theta +30\right)
\nonumber \\&& \quad\quad + 96 \cos^6\theta  \left[(4410 \xi -953) \cos^2\theta -18 (315 \xi -68)\right]  \left( M/r\right) + 4 (1848 \xi -367) \cos^8\theta   \left( M/r\right)^2
\\&&
A^{r r}{}_{4,2} \left[ \theta , M/r \right] =   -8 \cos^6\theta  \left[2 (2912 \xi -577) \cos^2\theta -7280 \xi +1429\right]  \left( M/r\right)^2
\\&&
A^{r r}{}_{4,3} \left[ \theta , M/r \right] = 0
\\&&
A^{r r}{}_{5,0} \left[ \theta , M/r \right] =   252 (32 \xi -7) \cos^8\theta  \left(3 \cos^2\theta -4\right)
\\&&
A^{r r}{}_{5,1} \left[ \theta , M/r \right] =   216 \cos^6\theta  \left(3 \cos^4\theta -12 \cos^2\theta +10\right)
\\&&
A^{r r}{}_{5,2} \left[ \theta , M/r \right] = 0
\\&&
A^{r r}{}_{5,3} \left[ \theta , M/r \right] = 0 .
\end{eqnarray}
\end{subequations}

For the massive scalar field, the coefficients $A^{\theta \theta}{}_{p,q}$ appearing in the expression (\ref{KN_SETthetatheta}) of $\langle T{}^\theta{}_\theta \rangle {}_\mathrm{ren}$ are

\begin{subequations}\label{KN_SETthetathetaScalar_coeff}
\begin{eqnarray}
&&
A^{\theta \theta}{}_{0,0} \left[ \theta , M/r \right] =   756 (32 \xi -7)  - 8 (7056 \xi -1543)  \left( M/r\right)
\\&&
A^{\theta \theta}{}_{0,1} \left[ \theta , M/r \right] =   -648  - 16 (4116 \xi -1093)  \left( M/r\right)+ 12 (15288 \xi -3649)  \left( M/r\right)^2
\\&&
A^{\theta \theta}{}_{0,2} \left[ \theta , M/r \right] =   16 (2912 \xi -739)  \left( M/r\right)^2- 16 (11928 \xi -2851)  \left( M/r\right)^3
\\&&
A^{\theta \theta}{}_{0,3} \left[ \theta , M/r \right] =   28 (2106 \xi -497)  \left( M/r\right)^4
\\&&
A^{\theta \theta}{}_{1,0} \left[ \theta , M/r \right] =   -252 (32 \xi -7) \left(85 \cos^2\theta -4\right)  + 288 (5516 \xi -1207) \cos^2\theta   \left( M/r\right)
\\&&
A^{\theta \theta}{}_{1,1} \left[ \theta , M/r \right] =   -648 \left(\cos^2\theta +4\right)  + 144 \left[(8204 \xi -1763) \cos^2\theta -588 \xi +169\right]  \left( M/r\right)
\nonumber \\&& \quad\quad - 16 (223104 \xi -48331) \cos^2\theta   \left( M/r\right)^2
\\&&
A^{\theta \theta}{}_{1,2} \left[ \theta , M/r \right] =   -8 \left[(55664 \xi -11807) \cos^2\theta -7280 \xi +1969\right]  \left( M/r\right)^2+ 112 (22048 \xi -4741) \cos^2\theta   \left( M/r\right)^3
\phantom{M/r}
\\&&
A^{\theta \theta}{}_{1,3} \left[ \theta , M/r \right] =   -4 (134414 \xi -28727) \cos^2\theta   \left( M/r\right)^4
\\&&
A^{\theta \theta}{}_{2,0} \left[ \theta , M/r \right] =   3528 (32 \xi -7) \cos^2\theta  \left(17 \cos^2\theta -8\right)  - 1008 (4088 \xi -895) \cos^4\theta   \left( M/r\right)
\\&&
A^{\theta \theta}{}_{2,1} \left[ \theta , M/r \right] =   432 \left(4 \cos^4\theta -14 \cos^2\theta -5\right)
\nonumber \\&& \quad\quad - 48 \cos^2\theta  \left[7 (5796 \xi -1261) \cos^2\theta -3 (10444 \xi -2181)\right]  \left( M/r\right) + 8 (766920 \xi -168349) \cos^4\theta   \left( M/r\right)^2
\\&&
A^{\theta \theta}{}_{2,2} \left[ \theta , M/r \right] =   8 \cos^2\theta  \left[7 (7152 \xi -1561) \cos^2\theta -68880 \xi +14029\right]  \left( M/r\right)^2
\nonumber \\&& \quad\quad - 272 (10136 \xi -2241) \cos^4\theta   \left( M/r\right)^3
\\&&
A^{\theta \theta}{}_{2,3} \left[ \theta , M/r \right] =   4 (88802 \xi -19981) \cos^4\theta   \left( M/r\right)^4
\\&&
A^{\theta \theta}{}_{3,0} \left[ \theta , M/r \right] =   -3528 (32 \xi -7) \cos^4\theta  \left(11 \cos^2\theta -20\right)  + 672 (2772 \xi -607) \cos^6\theta   \left( M/r\right)
\\&&
A^{\theta \theta}{}_{3,1} \left[ \theta , M/r \right] =   1296 \cos^2\theta  \left(2 \cos^4\theta -2 \cos^2\theta -5\right)
\nonumber \\&& \quad\quad + 16 \cos^4\theta  \left[(50820 \xi -11489) \cos^2\theta -9 (14980 \xi -3363)\right]  \left( M/r\right) - 16 (105168 \xi -23071) \cos^6\theta   \left( M/r\right)^2
\\&&
A^{\theta \theta}{}_{3,2} \left[ \theta , M/r \right] =   -8 \cos^4\theta  \left[(16016 \xi -3845) \cos^2\theta -50064 \xi +11899\right]  \left( M/r\right)^2
\nonumber \\&& \quad\quad + 16 (24304 \xi -5305) \cos^6\theta   \left( M/r\right)^3
\\&&
A^{\theta \theta}{}_{3,3} \left[ \theta , M/r \right] =   -4 (3962 \xi -773) \cos^6\theta   \left( M/r\right)^4
\\&&
A^{\theta \theta}{}_{4,0} \left[ \theta , M/r \right] =   252 (32 \xi -7) \cos^6\theta  \left(31 \cos^2\theta -112\right)  - 3528 (32 \xi -7) \cos^8\theta   \left( M/r\right)
\\&&
A^{\theta \theta}{}_{4,1} \left[ \theta , M/r \right] =   648 \cos^4\theta  \left(\cos^4\theta +4 \cos^2\theta -10\right)
\nonumber \\&& \quad\quad - 48 \cos^6\theta  \left[2 (1260 \xi -253) \cos^2\theta -3 (2660 \xi -559)\right]  \left( M/r\right) + 4 (10248 \xi -2267) \cos^8\theta   \left( M/r\right)^2
\\&&
A^{\theta \theta}{}_{4,2} \left[ \theta , M/r \right] =   8 \cos^6\theta  \left[(1456 \xi -167) \cos^2\theta -2800 \xi +327\right]  \left( M/r\right)^2
\\&&
A^{\theta \theta}{}_{4,3} \left[ \theta , M/r \right] = 0
\\&&
A^{\theta \theta}{}_{5,0} \left[ \theta , M/r \right] =   -252 (32 \xi -7) \cos^8\theta  \left(\cos^2\theta -4\right)
\\&&
A^{\theta \theta}{}_{5,1} \left[ \theta , M/r \right] =   -216 \cos^6\theta  \left(\cos^4\theta -8 \cos^2\theta +10\right)
\\&&
A^{\theta \theta}{}_{5,2} \left[ \theta , M/r \right] = 0
\\&&
A^{\theta \theta}{}_{5,3} \left[ \theta , M/r \right] = 0 .
\end{eqnarray}
\end{subequations}

For the massive scalar field, the coefficients $A^{\varphi \varphi}{}_{p,q}$ appearing in the expression (\ref{KN_SETphiphi}) of $\langle T{}^\varphi{}_\varphi \rangle {}_\mathrm{ren}$ are

\begin{subequations}\label{KN_SETphiphiScalar_coeff}
\begin{eqnarray}
&&
A^{\varphi \varphi}{}_{0,0} \left[ \theta , M/r \right] =   756 (32 \xi -7)  - 8 (7056 \xi -1543)  \left( M/r\right)
\\&&
A^{\varphi \varphi}{}_{0,1} \left[ \theta , M/r \right] =   -648  - 16 (4116 \xi -1093)  \left( M/r\right)+ 12 (15288 \xi -3649)  \left( M/r\right)^2
\\&&
A^{\varphi \varphi}{}_{0,2} \left[ \theta , M/r \right] =   16 (2912 \xi -739)  \left( M/r\right)^2- 16 (11928 \xi -2851)  \left( M/r\right)^3
\\&&
A^{\varphi \varphi}{}_{0,3} \left[ \theta , M/r \right] =   28 (2106 \xi -497)  \left( M/r\right)^4
\\&&
A^{\varphi \varphi}{}_{1,0} \left[ \theta , M/r \right] =   -20412 (32 \xi -7) \cos^2\theta   + 288 \left[10 (553 \xi -121) \cos^2\theta -14 \xi +3\right]  \left( M/r\right)
\\&&
A^{\varphi \varphi}{}_{1,1} \left[ \theta , M/r \right] =   216 \left(41 \cos^2\theta -56\right)  + 288 \left[(3920 \xi -953) \cos^2\theta -4 (28 \xi -39)\right]  \left( M/r\right)
\nonumber \\&& \quad\quad - 16 \left[9 (24892 \xi -5541) \cos^2\theta -2 (462 \xi -769)\right]  \left( M/r\right)^2
\\&&
A^{\varphi \varphi}{}_{1,2} \left[ \theta , M/r \right] =   -16 \left[2 (13216 \xi -3305) \cos^2\theta -2240 \xi +1691\right]  \left( M/r\right)^2
\nonumber \\&& \quad\quad + 16 \left[2 (77728 \xi -17373) \cos^2\theta -1120 \xi +1559\right]  \left( M/r\right)^3
\\&&
A^{\varphi \varphi}{}_{1,3} \left[ \theta , M/r \right] =   -4 \left[(136234 \xi -30411) \cos^2\theta -4 (455 \xi -421)\right]  \left( M/r\right)^4
\\&&
A^{\varphi \varphi}{}_{2,0} \left[ \theta , M/r \right] =   31752 (32 \xi -7) \cos^4\theta   - 144 \cos^2\theta  \left[(28868 \xi -6319) \cos^2\theta -18 (14 \xi -3)\right]  \left( M/r\right)
\\&&
A^{\varphi \varphi}{}_{2,1} \left[ \theta , M/r \right] =   432 \left(22 \cos^4\theta +38 \cos^2\theta -75\right)
\nonumber \\&& \quad\quad - 48 \left[13 (1176 \xi -205) \cos^4\theta -126 (48 \xi -11) \cos^2\theta -1005\right]  \left( M/r\right)
\nonumber \\&& \quad\quad + 8 \cos^2\theta  \left[(784560 \xi -167617) \cos^2\theta -12 (1470 \xi +61)\right]  \left( M/r\right)^2
\\&&
A^{\varphi \varphi}{}_{2,2} \left[ \theta , M/r \right] =   48 \left[ 414 \cos^4\theta -7 (448 \xi -87) \cos^2\theta -506\right]  \left( M/r\right)^2
\nonumber \\&& \quad\quad - 16 \cos^2\theta  \left[(181160 \xi -37343) \cos^2\theta -2 (4424 \xi +377)\right]  \left( M/r\right)^3
\\&&
A^{\varphi \varphi}{}_{2,3} \left[ \theta , M/r \right] =   4 \cos^2\theta  \left[(98602 \xi -19029) \cos^2\theta -56 (175 \xi +17)\right]  \left( M/r\right)^4
\\&&
A^{\varphi \varphi}{}_{3,0} \left[ \theta , M/r \right] =   31752 (32 \xi -7) \cos^6\theta   + 96 \cos^4\theta  \left[2 (9597 \xi -2102) \cos^2\theta +15 (14 \xi -3)\right]  \left( M/r\right)
\\&&
A^{\varphi \varphi}{}_{3,1} \left[ \theta , M/r \right] =   -432 \left(22 \cos^6\theta -132 \cos^4\theta +75 \cos^2\theta +50\right)
\nonumber \\&& \quad\quad - 32 \cos^2\theta  \left[2 (23520 \xi -5293) \cos^4\theta -18 (280 \xi -199) \cos^2\theta -2385\right]  \left( M/r\right)
\nonumber \\&& \quad\quad - 16 \cos^4\theta  \left[(108948 \xi -23677) \cos^2\theta -6 (630 \xi -101)\right]  \left( M/r\right)^2
\\&&
A^{\varphi \varphi}{}_{3,2} \left[ \theta , M/r \right] =   16 \cos^2\theta  \left[2 (13216 \xi -2927) \cos^4\theta -3 (3136 \xi -1509) \cos^2\theta -2700\right]  \left( M/r\right)^2
\nonumber \\&& \quad\quad + 16 \cos^4\theta  \left[28 (992 \xi -223) \cos^2\theta -3472 \xi +939\right]  \left( M/r\right)^3
\\&&
A^{\varphi \varphi}{}_{3,3} \left[ \theta , M/r \right] =   -4 \cos^4\theta  \left[7 (826 \xi -255) \cos^2\theta -4 (455 \xi -253)\right]  \left( M/r\right)^4
\\&&
A^{\varphi \varphi}{}_{4,0} \left[ \theta , M/r \right] =   -20412 (32 \xi -7) \cos^8\theta   - 72 \cos^6\theta  \left[(1288 \xi -283) \cos^2\theta +20 (14 \xi -3)\right]  \left( M/r\right)
\\&&
A^{\varphi \varphi}{}_{4,1} \left[ \theta , M/r \right] =   -216 \cos^2\theta  \left(41 \cos^6\theta -76 \cos^4\theta -150 \cos^2\theta +200\right)
\nonumber \\&& \quad\quad + 48 \cos^4\theta  \left[10 (882 \xi -169) \cos^4\theta -6 (560 \xi +11) \cos^2\theta +585\right]  \left( M/r\right)
\nonumber \\&& \quad\quad + 4 \cos^6\theta  \left[(9912 \xi -2131) \cos^2\theta +8 (42 \xi -17)\right]  \left( M/r\right)^2
\\&&
A^{\varphi \varphi}{}_{4,2} \left[ \theta , M/r \right] =   -16 \cos^4\theta  \left[(2912 \xi -253) \cos^4\theta -(2240 \xi +1009) \cos^2\theta +1182\right]  \left( M/r\right)^2
\\&&
A^{\varphi \varphi}{}_{4,3} \left[ \theta , M/r \right] = 0
\\&&
A^{\varphi \varphi}{}_{5,0} \left[ \theta , M/r \right] =   756 (32 \xi -7) \cos^{10}\theta
\\&&
A^{\varphi \varphi}{}_{5,1} \left[ \theta , M/r \right] =   216 \cos^4\theta  \left(\cos^2\theta -2\right) \left(3 \cos^4\theta -50 \cos^2\theta +50\right)
\\&&
A^{\varphi \varphi}{}_{5,2} \left[ \theta , M/r \right] = 0
\\&&
A^{\varphi \varphi}{}_{5,3} \left[ \theta , M/r \right] = 0 .
\end{eqnarray}
\end{subequations}

For the massive scalar field, the coefficients $A^{t \varphi}{}_{p,q}$ appearing in the expression (\ref{KN_SETtphi}) of $\langle T{}^t{}_\varphi \rangle {}_\mathrm{ren}$ are

\begin{subequations}\label{KN_SETtphiScalar_coeff}
\begin{eqnarray}
&&
A^{t \varphi}{}_{0,0} \left[ \theta , M/r \right] =   -72 (84 \xi -17)  \left( M/r\right)
\\&&
A^{t \varphi}{}_{0,1} \left[ \theta , M/r \right] =   -2160  + 7056  \left( M/r\right)+ 28 (672 \xi -293)  \left( M/r\right)^2
\\&&
A^{t \varphi}{}_{0,2} \left[ \theta , M/r \right] =   -3444  \left( M/r\right)^2- 32 (609 \xi -253)  \left( M/r\right)^3
\\&&
A^{t \varphi}{}_{0,3} \left[ \theta , M/r \right] =   72 (91 \xi -32)  \left( M/r\right)^4
\\&&
A^{t \varphi}{}_{1,0} \left[ \theta , M/r \right] =   144 \left[(910 \xi -181) \cos^2\theta -14 \xi +3\right]  \left( M/r\right)
\\&&
A^{t \varphi}{}_{1,1} \left[ \theta , M/r \right] =   2160 \left(\cos^2\theta -6\right)  - 72 \left(33 \cos^2\theta -397\right)  \left( M/r\right)
\nonumber \\&& \quad\quad - 16 \left[(18606 \xi -3505) \cos^2\theta -462 \xi +769\right]  \left( M/r\right)^2
\\&&
A^{t \varphi}{}_{1,2} \left[ \theta , M/r \right] =   -12 \left(13 \cos^2\theta +1191\right)  \left( M/r\right)^2+ 8 \left[7 (3836 \xi -677) \cos^2\theta -1120 \xi +1559\right]  \left( M/r\right)^3
\\&&
A^{t \varphi}{}_{1,3} \left[ \theta , M/r \right] =   -8 \left[(6118 \xi -1011) \cos^2\theta -455 \xi +421\right]  \left( M/r\right)^4
\\&&
A^{t \varphi}{}_{2,0} \left[ \theta , M/r \right] =   -144 \cos^2\theta  \left[10 (189 \xi -37) \cos^2\theta -9 (14 \xi -3)\right]  \left( M/r\right)
\\&&
A^{t \varphi}{}_{2,1} \left[ \theta , M/r \right] =   4320 \left(2 \cos^4\theta -2 \cos^2\theta -5\right)  - 72 \left(240 \cos^4\theta -353 \cos^2\theta -335\right)  \left( M/r\right)
\nonumber \\&& \quad\quad + 8 \cos^2\theta  \left[(55860 \xi -9617) \cos^2\theta -6 (1470 \xi +61)\right]  \left( M/r\right)^2
\\&&
A^{t \varphi}{}_{2,2} \left[ \theta , M/r \right] =   12 \left(671 \cos^4\theta -1381 \cos^2\theta -1012\right)  \left( M/r\right)^2
\nonumber \\&& \quad\quad - 16 \cos^2\theta  \left[(14014 \xi -2133) \cos^2\theta -4424 \xi -377\right]  \left( M/r\right)^3
\\&&
A^{t \varphi}{}_{2,3} \left[ \theta , M/r \right] =   8 \cos^2\theta  \left[(4123 \xi -530) \cos^2\theta -14 (175 \xi +17)\right]  \left( M/r\right)^4
\\&&
A^{t \varphi}{}_{3,0} \left[ \theta , M/r \right] =   144 \cos^4\theta  \left[(714 \xi -139) \cos^2\theta +5 (14 \xi -3)\right]  \left( M/r\right)
\\&&
A^{t \varphi}{}_{3,1} \left[ \theta , M/r \right] =   2160 \left(\cos^6\theta +9 \cos^4\theta -15 \cos^2\theta -5\right)  - 72 \cos^2\theta  \left(73 \cos^4\theta +261 \cos^2\theta -530\right)  \left( M/r\right)
\nonumber \\&& \quad\quad - 16 \cos^4\theta  \left[(7266 \xi -1525) \cos^2\theta -3 (630 \xi -101)\right]  \left( M/r\right)^2
\\&&
A^{t \varphi}{}_{3,2} \left[ \theta , M/r \right] =   12 \cos^2\theta  \left(289 \cos^4\theta +531 \cos^2\theta -1800\right)  \left( M/r\right)^2
\nonumber \\&& \quad\quad + 8 \cos^4\theta  \left[(4508 \xi -1231) \cos^2\theta -3472 \xi +939\right]  \left( M/r\right)^3
\\&&
A^{t \varphi}{}_{3,3} \left[ \theta , M/r \right] =   -8 \cos^4\theta  \left[(364 \xi -191) \cos^2\theta -455 \xi +253\right]  \left( M/r\right)^4
\\&&
A^{t \varphi}{}_{4,0} \left[ \theta , M/r \right] =   -72 \cos^6\theta  \left[(56 \xi -11) \cos^2\theta +10 (14 \xi -3)\right]  \left( M/r\right)
\\&&
A^{t \varphi}{}_{4,1} \left[ \theta , M/r \right] =   -2160 \cos^2\theta  \left(\cos^6\theta -6 \cos^4\theta +10\right)  + 72 \cos^4\theta  \left(36 \cos^4\theta -217 \cos^2\theta +195\right)  \left( M/r\right)
\nonumber \\&& \quad\quad + 4 \cos^6\theta  \left[9 (56 \xi -11) \cos^2\theta +4 (42 \xi -17)\right]  \left( M/r\right)^2
\\&&
A^{t \varphi}{}_{4,2} \left[ \theta , M/r \right] =   -12 \cos^4\theta  \left(108 \cos^4\theta -721 \cos^2\theta +788\right)  \left( M/r\right)^2
\\&&
A^{t \varphi}{}_{4,3} \left[ \theta , M/r \right] = 0
\\&&
A^{t \varphi}{}_{5,0} \left[ \theta , M/r \right] = 0
\\&&
A^{t \varphi}{}_{5,1} \left[ \theta , M/r \right] =   -2160 \cos^4\theta  \left(\cos^4\theta -5 \cos^2\theta +5\right)
\\&&
A^{t \varphi}{}_{5,2} \left[ \theta , M/r \right] = 0
\\&&
A^{t \varphi}{}_{5,3} \left[ \theta , M/r \right] = 0 .
\end{eqnarray}
\end{subequations}

For the massive scalar field, the coefficients $A^{\varphi t}{}_{p,q}$ appearing in the expression (\ref{KN_SETphit}) of $\langle T{}^\varphi{}_t \rangle {}_\mathrm{ren}$ are

\begin{subequations}\label{KN_SETphitScalar_coeff}
\begin{eqnarray}
&&
A^{\varphi t}{}_{0,0} \left[ \theta , M/r \right] =   144 (14 \xi -3)  \left( M/r\right)
\\&&
A^{\varphi t}{}_{0,1} \left[ \theta , M/r \right] =   2160  - 9648  \left( M/r\right)- 16 (462 \xi -769)  \left( M/r\right)^2
\\&&
A^{\varphi t}{}_{0,2} \left[ \theta , M/r \right] =   4740  \left( M/r\right)^2+ 8 (1120 \xi -1559)  \left( M/r\right)^3
\\&&
A^{\varphi t}{}_{0,3} \left[ \theta , M/r \right] =   -8 (455 \xi -421)  \left( M/r\right)^4
\\&&
A^{\varphi t}{}_{1,0} \left[ \theta , M/r \right] =   -1296 (14 \xi -3) \cos^2\theta   \left( M/r\right)
\\&&
A^{\varphi t}{}_{1,1} \left[ \theta , M/r \right] =   -2160 \left(\cos^2\theta -5\right)  - 72 \left(39 \cos^2\theta +335\right)  \left( M/r\right)+ 48 (1470 \xi +61) \cos^2\theta   \left( M/r\right)^2
\\&&
A^{\varphi t}{}_{1,2} \left[ \theta , M/r \right] =   12 \left(229 \cos^2\theta +1012\right)  \left( M/r\right)^2- 16 (4424 \xi +377) \cos^2\theta   \left( M/r\right)^3
\\&&
A^{\varphi t}{}_{1,3} \left[ \theta , M/r \right] =   112 (175 \xi +17) \cos^2\theta   \left( M/r\right)^4
\\&&
A^{\varphi t}{}_{2,0} \left[ \theta , M/r \right] =   -720 (14 \xi -3) \cos^4\theta   \left( M/r\right)
\\&&
A^{\varphi t}{}_{2,1} \left[ \theta , M/r \right] =   -2160 \left(4 \cos^4\theta -5 \cos^2\theta -5\right)  + 720 \cos^2\theta  \left(24 \cos^2\theta -53\right)  \left( M/r\right)
\nonumber \\&& \quad\quad - 48 (630 \xi -101) \cos^4\theta   \left( M/r\right)^2
\\&&
A^{\varphi t}{}_{2,2} \left[ \theta , M/r \right] =   -12 \cos^2\theta  \left(671 \cos^2\theta -1800\right)  \left( M/r\right)^2+ 8 (3472 \xi -939) \cos^4\theta   \left( M/r\right)^3
\\&&
A^{\varphi t}{}_{2,3} \left[ \theta , M/r \right] =   -8 (455 \xi -253) \cos^4\theta   \left( M/r\right)^4
\\&&
A^{\varphi t}{}_{3,0} \left[ \theta , M/r \right] =   720 (14 \xi -3) \cos^6\theta   \left( M/r\right)
\\&&
A^{\varphi t}{}_{3,1} \left[ \theta , M/r \right] =   -2160 \cos^2\theta  \left(\cos^4\theta +5 \cos^2\theta -10\right)  + 360 \cos^4\theta  \left(29 \cos^2\theta -39\right)  \left( M/r\right)
\nonumber \\&& \quad\quad - 16 (42 \xi -17) \cos^6\theta   \left( M/r\right)^2
\\&&
A^{\varphi t}{}_{3,2} \left[ \theta , M/r \right] =   -12 \cos^4\theta  \left(505 \cos^2\theta -788\right)  \left( M/r\right)^2
\\&&
A^{\varphi t}{}_{3,3} \left[ \theta , M/r \right] = 0
\\&&
A^{\varphi t}{}_{4,0} \left[ \theta , M/r \right] = 0
\\&&
A^{\varphi t}{}_{4,1} \left[ \theta , M/r \right] =   2160 \cos^4\theta  \left(\cos^4\theta -5 \cos^2\theta +5\right)
\\&&
A^{\varphi t}{}_{4,2} \left[ \theta , M/r \right] = 0
\\&&
A^{\varphi t}{}_{4,3} \left[ \theta , M/r \right] = 0 .
\end{eqnarray}
\end{subequations}

For the massive scalar field, the coefficients $A^{r \theta}{}_{p,q}$ appearing in the expression (\ref{KN_SETrtheta}) of $\langle T{}^r{}_\theta \rangle {}_\mathrm{ren}$ are

\begin{subequations}\label{KN_SETrthetaScalar_coeff}
\begin{eqnarray}
&&
A^{r \theta}{}_{0,0} \left[ \theta , M/r \right] =   72 (32 \xi -7)  - 144 (32 \xi -7)  \left( M/r\right)
\\&&
A^{r \theta}{}_{0,1} \left[ \theta , M/r \right] =   -36 (109 \xi -24)  \left( M/r\right)+ 72 (141 \xi -31)  \left( M/r\right)^2
\\&&
A^{r \theta}{}_{0,2} \left[ \theta , M/r \right] =   10 (168 \xi -37)  \left( M/r\right)^2- 4 (1821 \xi -401)  \left( M/r\right)^3
\\&&
A^{r \theta}{}_{0,3} \left[ \theta , M/r \right] =   10 (168 \xi -37)  \left( M/r\right)^4
\\&&
A^{r \theta}{}_{1,0} \left[ \theta , M/r \right] =   -72 (32 \xi -7) \left(7 \cos^2\theta -1\right)  + 1008 (32 \xi -7) \cos^2\theta   \left( M/r\right)
\\&&
A^{r \theta}{}_{1,1} \left[ \theta , M/r \right] =   12 \left[5 (339 \xi -74) \cos^2\theta -3 (109 \xi -24)\right]  \left( M/r\right)- 24 (2367 \xi -517) \cos^2\theta   \left( M/r\right)^2
\\&&
A^{r \theta}{}_{1,2} \left[ \theta , M/r \right] =   -2 \left[2 (1464 \xi -319) \cos^2\theta -5 (168 \xi -37)\right]  \left( M/r\right)^2+ 4 (8013 \xi -1748) \cos^2\theta   \left( M/r\right)^3
\\&&
A^{r \theta}{}_{1,3} \left[ \theta , M/r \right] =   -4 (1464 \xi -319) \cos^2\theta   \left( M/r\right)^4
\\&&
A^{r \theta}{}_{2,0} \left[ \theta , M/r \right] =   504 (32 \xi -7) \cos^2\theta  (\cos^2\theta -1)  - 1008 (32 \xi -7) \cos^4\theta   \left( M/r\right)
\\&&
A^{r \theta}{}_{2,1} \left[ \theta , M/r \right] =   -60 \cos^2\theta  \left[(201 \xi -44) \cos^2\theta -339 \xi +74\right]  \left( M/r\right)+ 24 (1677 \xi -367) \cos^4\theta   \left( M/r\right)^2
\\&&
A^{r \theta}{}_{2,2} \left[ \theta , M/r \right] =   2 \cos^2\theta  \left[5 (168 \xi -37) \cos^2\theta -2 (1464 \xi -319)\right]  \left( M/r\right)^2- 20 (771 \xi -169) \cos^4\theta   \left( M/r\right)^3
\\&&
A^{r \theta}{}_{2,3} \left[ \theta , M/r \right] =   10 (168 \xi -37) \cos^4\theta   \left( M/r\right)^4
\\&&
A^{r \theta}{}_{3,0} \left[ \theta , M/r \right] =   -72 (32 \xi -7) \cos^4\theta  \left(\cos^2\theta -7\right)  + 144 (32 \xi -7) \cos^6\theta   \left( M/r\right)
\\&&
A^{r \theta}{}_{3,1} \left[ \theta , M/r \right] =   60 \cos^4\theta  \left[(9 \xi -2) \cos^2\theta -201 \xi +44\right]  \left( M/r\right)- 24 (141 \xi -31) \cos^6\theta   \left( M/r\right)^2
\\&&
A^{r \theta}{}_{3,2} \left[ \theta , M/r \right] =   10 (168 \xi -37) \cos^4\theta   \left( M/r\right)^2+ 60 (9 \xi -2) \cos^6\theta   \left( M/r\right)^3
\\&&
A^{r \theta}{}_{3,3} \left[ \theta , M/r \right] = 0
\\&&
A^{r \theta}{}_{4,0} \left[ \theta , M/r \right] =   -72 (32 \xi -7) \cos^6\theta
\\&&
A^{r \theta}{}_{4,1} \left[ \theta , M/r \right] =   60 (9 \xi -2) \cos^6\theta   \left( M/r\right)
\\&&
A^{r \theta}{}_{4,2} \left[ \theta , M/r \right] = 0
\\&&
A^{r \theta}{}_{4,3} \left[ \theta , M/r \right] = 0 .
\end{eqnarray}
\end{subequations}

For the massive scalar field, the coefficients $A^{\theta r}{}_{p,q}$ appearing in the expression (\ref{KN_SETthetar}) of $\langle T{}^\theta{}_r \rangle {}_\mathrm{ren}$ are

\begin{subequations}\label{KN_SETthetarScalar_coeff}
\begin{eqnarray}
&&
A^{\theta r}{}_{0,0} \left[ \theta , M/r \right] =   72 (32 \xi -7)
\\&&
A^{\theta r}{}_{0,1} \left[ \theta , M/r \right] =   -36 (109 \xi -24)  \left( M/r\right)
\\&&
A^{\theta r}{}_{0,2} \left[ \theta , M/r \right] =   10 (168 \xi -37)  \left( M/r\right)^2
\\&&
A^{\theta r}{}_{1,0} \left[ \theta , M/r \right] =   -504 (32 \xi -7) \cos^2\theta
\\&&
A^{\theta r}{}_{1,1} \left[ \theta , M/r \right] =   60 (339 \xi -74) \cos^2\theta   \left( M/r\right)
\\&&
A^{\theta r}{}_{1,2} \left[ \theta , M/r \right] =   -4 (1464 \xi -319) \cos^2\theta   \left( M/r\right)^2
\\&&
A^{\theta r}{}_{2,0} \left[ \theta , M/r \right] =   504 (32 \xi -7) \cos^4\theta
\\&&
A^{\theta r}{}_{2,1} \left[ \theta , M/r \right] =   -60 (201 \xi -44) \cos^4\theta   \left( M/r\right)
\\&&
A^{\theta r}{}_{2,2} \left[ \theta , M/r \right] =   10 (168 \xi -37) \cos^4\theta   \left( M/r\right)^2
\\&&
A^{\theta r}{}_{3,0} \left[ \theta , M/r \right] =   -72 (32 \xi -7) \cos^6\theta
\\&&
A^{\theta r}{}_{3,1} \left[ \theta , M/r \right] =   60 (9 \xi -2) \cos^6\theta   \left( M/r\right)
\\&&
A^{\theta r}{}_{3,2} \left[ \theta , M/r \right] = 0 .
\end{eqnarray}
\end{subequations}

\subsection{Massive Dirac field} \label{KN_SETDirac_coeff}

For the massive Dirac field, the coefficients $A^{t t}{}_{p,q}$ appearing in the expression (\ref{KN_SETtt}) of $\langle T{}^t{}_t \rangle {}_\mathrm{ren}$ are

\begin{subequations}\label{KN_SETttDirac_coeff}
\begin{eqnarray}
&&
A^{t t}{}_{0,0} \left[ \theta , M/r \right] =   -1080  + 2384  \left( M/r\right)
\\&&
A^{t t}{}_{0,1} \left[ \theta , M/r \right] =   5400  - 22464  \left( M/r\right)+ 21832  \left( M/r\right)^2
\\&&
A^{t t}{}_{0,2} \left[ \theta , M/r \right] =   10544  \left( M/r\right)^2- 21496  \left( M/r\right)^3
\\&&
A^{t t}{}_{0,3} \left[ \theta , M/r \right] =   4917  \left( M/r\right)^4
\\&&
A^{t t}{}_{1,0} \left[ \theta , M/r \right] =   29160 \cos^2\theta   - 288 \left(251 \cos^2\theta +1\right)  \left( M/r\right)
\\&&
A^{t t}{}_{1,1} \left[ \theta , M/r \right] =   -5400 \left(7 \cos^2\theta -12\right)  + 288 \left(185 \cos^2\theta -691\right)  \left( M/r\right)+ 16 \left(7805 \cos^2\theta +7277\right)  \left( M/r\right)^2
\\&&
A^{t t}{}_{1,2} \left[ \theta , M/r \right] =   -64 \left(497 \cos^2\theta -1616\right)  \left( M/r\right)^2- 8 \left(10010 \cos^2\theta +15373\right)  \left( M/r\right)^3
\\&&
A^{t t}{}_{1,3} \left[ \theta , M/r \right] =  (17765 \cos^2\theta +32546) \left( M/r\right)^4
\\&&
A^{t t}{}_{2,0} \left[ \theta , M/r \right] =   -45360 \cos^4\theta   + 2592 \cos^2\theta  \left(76 \cos^2\theta +1\right)  \left( M/r\right)
\\&&
A^{t t}{}_{2,1} \left[ \theta , M/r \right] =   -10800 \left(4 \cos^4\theta +6 \cos^2\theta -15\right)  + 288 \left(628 \cos^4\theta -111 \cos^2\theta -855\right)  \left( M/r\right)
\nonumber \\&& \quad\quad - 240 \cos^2\theta  \left(1658 \cos^2\theta -535\right)  \left( M/r\right)^2
\\&&
A^{t t}{}_{2,2} \left[ \theta , M/r \right] =   -96 \left(900 \cos^4\theta -422 \cos^2\theta -1349\right)  \left( M/r\right)^2+ 8 \cos^2\theta  \left(30793 \cos^2\theta -20378\right)  \left( M/r\right)^3
\\&&
A^{t t}{}_{2,3} \left[ \theta , M/r \right] =   -\cos^2\theta  \left(48149 \cos^2\theta -49476\right)  \left( M/r\right)^4
\\&&
A^{t t}{}_{3,0} \left[ \theta , M/r \right] =   -45360 \cos^6\theta   - 96 \cos^4\theta  \left(967 \cos^2\theta -15\right)  \left( M/r\right)
\\&&
A^{t t}{}_{3,1} \left[ \theta , M/r \right] =   10800 \left(4 \cos^6\theta -24 \cos^4\theta +15 \cos^2\theta +10\right)
\nonumber \\&& \quad\quad + 1440 \cos^2\theta  \left(21 \cos^4\theta +247 \cos^2\theta -258\right)  \left( M/r\right) + 16 \cos^4\theta  \left(4751 \cos^2\theta -645\right)  \left( M/r\right)^2
\\&&
A^{t t}{}_{3,2} \left[ \theta , M/r \right] =   -64 \cos^2\theta  \left(178 \cos^4\theta +2742 \cos^2\theta -3375\right)  \left( M/r\right)^2+ 24 \cos^4\theta  \left(140 \cos^2\theta -687\right)  \left( M/r\right)^3
\phantom{M}
\\&&
A^{t t}{}_{3,3} \left[ \theta , M/r \right] =   -\cos^4\theta  \left(8773 \cos^2\theta -11042\right)  \left( M/r\right)^4
\\&&
A^{t t}{}_{4,0} \left[ \theta , M/r \right] =   29160 \cos^8\theta   + 144 \cos^6\theta  \left(43 \cos^2\theta -10\right)  \left( M/r\right)
\\&&
A^{t t}{}_{4,1} \left[ \theta , M/r \right] =   5400 \cos^2\theta  \left(7 \cos^6\theta -12 \cos^4\theta -30 \cos^2\theta +40\right)
\nonumber \\&& \quad\quad - 1440 \cos^4\theta  \left(52 \cos^4\theta -131 \cos^2\theta +87\right)  \left( M/r\right) - 8 \cos^6\theta  \left(325 \cos^2\theta -158\right)  \left( M/r\right)^2
\\&&
A^{t t}{}_{4,2} \left[ \theta , M/r \right] =   16 \cos^4\theta  \left(2041 \cos^4\theta -7036 \cos^2\theta +5406\right)  \left( M/r\right)^2
\\&&
A^{t t}{}_{4,3}\left[ \theta , M/r \right] = 0
\\&&
A^{t t}{}_{5,0} \left[ \theta , M/r \right] =   -1080 \cos^{10}\theta
\\&&
A^{t t}{}_{5,1} \left[ \theta , M/r \right] =   -5400 \cos^4\theta  \left(\cos^2\theta -2\right) \left(\cos^4\theta -10 \cos^2\theta +10\right)
\\&&
A^{t t}{}_{5,2}\left[ \theta , M/r \right] = 0
\\&&
A^{t t}{}_{5,3}\left[ \theta , M/r \right] = 0 .
\end{eqnarray}
\end{subequations}

For the massive Dirac field, the coefficients $A^{r r}{}_{p,q}$ appearing in the expression (\ref{KN_SETrr}) of $\langle T{}^r{}_r \rangle {}_\mathrm{ren}$ are

\begin{subequations}\label{KN_SETrrDirac_coeff}
\begin{eqnarray}
&&
A^{r r}{}_{0,0} \left[ \theta , M/r \right] =   504  - 784  \left( M/r\right)
\\&&
A^{r r}{}_{0,1} \left[ \theta , M/r \right] =   1080  - 6336  \left( M/r\right)+ 8440  \left( M/r\right)^2
\\&&
A^{r r}{}_{0,2} \left[ \theta , M/r \right] =   3560  \left( M/r\right)^2- 8680  \left( M/r\right)^3
\\&&
A^{r r}{}_{0,3} \left[ \theta , M/r \right] =   2253  \left( M/r\right)^4
\\&&
A^{r r}{}_{1,0} \left[ \theta , M/r \right] =   -504 \left(31 \cos^2\theta -4\right)  + 16128 \cos^2\theta   \left( M/r\right)
\\&&
A^{r r}{}_{1,1} \left[ \theta , M/r \right] =   -1080 \left(3 \cos^2\theta -8\right)  + 144 \left(189 \cos^2\theta -169\right)  \left( M/r\right)- 16160 \cos^2\theta   \left( M/r\right)^2
\\&&
A^{r r}{}_{1,2} \left[ \theta , M/r \right] =   -8 \left(1141 \cos^2\theta -1563\right)  \left( M/r\right)^2+ 88 \cos^2\theta   \left( M/r\right)^3
\\&&
A^{r r}{}_{1,3} \left[ \theta , M/r \right] =   2015 \cos^2\theta   \left( M/r\right)^4
\\&&
A^{r r}{}_{2,0} \left[ \theta , M/r \right] =   7056 \cos^2\theta  \left(11 \cos^2\theta -8\right)  - 22176 \cos^4\theta   \left( M/r\right)
\\&&
A^{r r}{}_{2,1} \left[ \theta , M/r \right] =   -2160 \left(6 \cos^4\theta -6 \cos^2\theta -5\right)  - 144 \cos^2\theta  \left(429 \cos^2\theta -193\right)  \left( M/r\right)
\nonumber \\&& \quad\quad + 31440 \cos^4\theta   \left( M/r\right)^2
\\&&
A^{r r}{}_{2,2} \left[ \theta , M/r \right] =   8 \cos^2\theta  \left(1313 \cos^2\theta +307\right)  \left( M/r\right)^2- 18840 \cos^4\theta   \left( M/r\right)^3
\\&&
A^{r r}{}_{2,3} \left[ \theta , M/r \right] =   4887 \cos^4\theta   \left( M/r\right)^4
\\&&
A^{r r}{}_{3,0} \left[ \theta , M/r \right] =   -7056 \cos^4\theta  \left(17 \cos^2\theta -20\right)  - 2688 \cos^6\theta   \left( M/r\right)
\\&&
A^{r r}{}_{3,1} \left[ \theta , M/r \right] =   -2160 \cos^2\theta  \left(4 \cos^4\theta +6 \cos^2\theta -15\right)  + 9360 \cos^4\theta  \left(15 \cos^2\theta -19\right)  \left( M/r\right)
\nonumber \\&& \quad\quad + 4064 \cos^6\theta   \left( M/r\right)^2
\\&&
A^{r r}{}_{3,2} \left[ \theta , M/r \right] =   -8 \cos^4\theta  \left(5275 \cos^2\theta -7013\right)  \left( M/r\right)^2- 1496 \cos^6\theta   \left( M/r\right)^3
\\&&
A^{r r}{}_{3,3} \left[ \theta , M/r \right] =   773 \cos^6\theta   \left( M/r\right)^4
\\&&
A^{r r}{}_{4,0} \left[ \theta , M/r \right] =   504 \cos^6\theta  \left(85 \cos^2\theta -112\right)  + 1584 \cos^8\theta   \left( M/r\right)
\\&&
A^{r r}{}_{4,1} \left[ \theta , M/r \right] =   1080 \cos^4\theta  \left(3 \cos^4\theta -28 \cos^2\theta +30\right)  - 720 \cos^6\theta  \left(31 \cos^2\theta -39\right)  \left( M/r\right)
\nonumber \\&& \quad\quad + 248 \cos^8\theta   \left( M/r\right)^2
\\&&
A^{r r}{}_{4,2} \left[ \theta , M/r \right] =   -40 \cos^6\theta  \left(22 \cos^2\theta -41\right)  \left( M/r\right)^2
\\&&
A^{r r}{}_{4,3}\left[ \theta , M/r \right] = 0
\\&&
A^{r r}{}_{5,0} \left[ \theta , M/r \right] =   -504 \cos^8\theta  \left(3 \cos^2\theta -4\right)
\\&&
A^{r r}{}_{5,1} \left[ \theta , M/r \right] =   1080 \cos^6\theta  \left(3 \cos^4\theta -12 \cos^2\theta +10\right)
\\&&
A^{r r}{}_{5,2}\left[ \theta , M/r \right] = 0
\\&&
A^{r r}{}_{5,3}\left[ \theta , M/r \right] = 0 .
\end{eqnarray}
\end{subequations}

For the massive Dirac field, the coefficients $A^{\theta \theta}{}_{p,q}$ appearing in the expression (\ref{KN_SETthetatheta}) of $\langle T{}^\theta{}_\theta \rangle {}_\mathrm{ren}$ are

\begin{subequations}\label{KN_SETthetathetaDirac_coeff}
\begin{eqnarray}
&&
A^{\theta \theta}{}_{0,0} \left[ \theta , M/r \right] =   -1512  + 3536  \left( M/r\right)
\\&&
A^{\theta \theta}{}_{0,1} \left[ \theta , M/r \right] =   -3240  + 20016  \left( M/r\right)- 30808  \left( M/r\right)^2
\\&&
A^{\theta \theta}{}_{0,2} \left[ \theta , M/r \right] =   -12080  \left( M/r\right)^2+ 33984  \left( M/r\right)^3
\\&&
A^{\theta \theta}{}_{0,3} \left[ \theta , M/r \right] =   -9933  \left( M/r\right)^4
\\&&
A^{\theta \theta}{}_{1,0} \left[ \theta , M/r \right] =   504 \left(85 \cos^2\theta -4\right)  - 99072 \cos^2\theta   \left( M/r\right)
\\&&
A^{\theta \theta}{}_{1,1} \left[ \theta , M/r \right] =   -3240 \left(\cos^2\theta +4\right)  - 144 \left(359 \cos^2\theta -243\right)  \left( M/r\right)+ 186272 \cos^2\theta   \left( M/r\right)^2
\\&&
A^{\theta \theta}{}_{1,2} \left[ \theta , M/r \right] =   40 \left(299 \cos^2\theta -499\right)  \left( M/r\right)^2- 106960 \cos^2\theta   \left( M/r\right)^3
\\&&
A^{\theta \theta}{}_{1,3} \left[ \theta , M/r \right] =   18817 \cos^2\theta   \left( M/r\right)^4
\\&&
A^{\theta \theta}{}_{2,0} \left[ \theta , M/r \right] =   -7056 \cos^2\theta  \left(17 \cos^2\theta -8\right)  + 256032 \cos^4\theta   \left( M/r\right)
\\&&
A^{\theta \theta}{}_{2,1} \left[ \theta , M/r \right] =   2160 \left(4 \cos^4\theta -14 \cos^2\theta -5\right)  + 144 \cos^2\theta  \left(749 \cos^2\theta -139\right)  \left( M/r\right)
\nonumber \\&& \quad\quad - 393360 \cos^4\theta   \left( M/r\right)^2
\\&&
A^{\theta \theta}{}_{2,2} \left[ \theta , M/r \right] =   -8 \cos^2\theta  \left(2933 \cos^2\theta +1087\right)  \left( M/r\right)^2+ 196128 \cos^4\theta   \left( M/r\right)^3
\\&&
A^{\theta \theta}{}_{2,3} \left[ \theta , M/r \right] =   -31959 \cos^4\theta   \left( M/r\right)^4
\\&&
A^{\theta \theta}{}_{3,0} \left[ \theta , M/r \right] =   7056 \cos^4\theta  \left(11 \cos^2\theta -20\right)  - 115584 \cos^6\theta   \left( M/r\right)
\\&&
A^{\theta \theta}{}_{3,1} \left[ \theta , M/r \right] =   6480 \cos^2\theta  \left(2 \cos^4\theta -2 \cos^2\theta -5\right)  - 720 \cos^4\theta  \left(113 \cos^2\theta -269\right)  \left( M/r\right)
\nonumber \\&& \quad\quad + 108448 \cos^6\theta   \left( M/r\right)^2
\\&&
A^{\theta \theta}{}_{3,2} \left[ \theta , M/r \right] =   8 \cos^4\theta  \left(2761 \cos^2\theta -7793\right)  \left( M/r\right)^2- 23888 \cos^6\theta   \left( M/r\right)^3
\\&&
A^{\theta \theta}{}_{3,3} \left[ \theta , M/r \right] =   -549 \cos^6\theta   \left( M/r\right)^4
\\&&
A^{\theta \theta}{}_{4,0} \left[ \theta , M/r \right] =   -504 \cos^6\theta  \left(31 \cos^2\theta -112\right)  + 7056 \cos^8\theta   \left( M/r\right)
\\&&
A^{\theta \theta}{}_{4,1} \left[ \theta , M/r \right] =   3240 \cos^4\theta  \left(\cos^4\theta +4 \cos^2\theta -10\right)  - 720 \cos^6\theta  \left(4 \cos^2\theta +13\right)  \left( M/r\right)
\nonumber \\&& \quad\quad - 3032 \cos^8\theta   \left( M/r\right)^2
\\&&
A^{\theta \theta}{}_{4,2} \left[ \theta , M/r \right] =   8 \cos^6\theta  \left(635 \cos^2\theta -1137\right)  \left( M/r\right)^2
\\&&
A^{\theta \theta}{}_{4,3}\left[ \theta , M/r \right] = 0
\\&&
A^{\theta \theta}{}_{5,0} \left[ \theta , M/r \right] =   504 \cos^8\theta  \left(\cos^2\theta -4\right)
\\&&
A^{\theta \theta}{}_{5,1} \left[ \theta , M/r \right] =   -1080 \cos^6\theta  \left(\cos^4\theta -8 \cos^2\theta +10\right)
\\&&
A^{\theta \theta}{}_{5,2}\left[ \theta , M/r \right] = 0
\\&&
A^{\theta \theta}{}_{5,3}\left[ \theta , M/r \right] = 0 .
\end{eqnarray}
\end{subequations}

For the massive Dirac field, the coefficients $A^{\varphi \varphi}{}_{p,q}$ appearing in the expression (\ref{KN_SETphiphi}) of $\langle T{}^\varphi{}_\varphi \rangle {}_\mathrm{ren}$ are

\begin{subequations}\label{KN_SETphiphiDirac_coeff}
\begin{eqnarray}
&&
A^{\varphi \varphi}{}_{0,0} \left[ \theta , M/r \right] =   -1512  + 3536  \left( M/r\right)
\\&&
A^{\varphi \varphi}{}_{0,1} \left[ \theta , M/r \right] =   -3240  + 20016  \left( M/r\right)- 30808  \left( M/r\right)^2
\\&&
A^{\varphi \varphi}{}_{0,2} \left[ \theta , M/r \right] =   -12080  \left( M/r\right)^2+ 33984  \left( M/r\right)^3
\\&&
A^{\varphi \varphi}{}_{0,3} \left[ \theta , M/r \right] =   -9933  \left( M/r\right)^4
\\&&
A^{\varphi \varphi}{}_{1,0} \left[ \theta , M/r \right] =   40824 \cos^2\theta   - 288 \left(345 \cos^2\theta -1\right)  \left( M/r\right)
\\&&
A^{\varphi \varphi}{}_{1,1} \left[ \theta , M/r \right] =   1080 \left(41 \cos^2\theta -56\right)  - 576 \left(370 \cos^2\theta -341\right)  \left( M/r\right)+ 16 \left(18919 \cos^2\theta -7277\right)  \left( M/r\right)^2
\phantom{M/r}
\\&&
A^{\varphi \varphi}{}_{1,2} \left[ \theta , M/r \right] =   160 \left(617 \cos^2\theta -667\right)  \left( M/r\right)^2- 8 \left(28743 \cos^2\theta -15373\right)  \left( M/r\right)^3
\\&&
A^{\varphi \varphi}{}_{1,3} \left[ \theta , M/r \right] =  (51363 \cos^2\theta -32546) \left( M/r\right)^4
\\&&
A^{\varphi \varphi}{}_{2,0} \left[ \theta , M/r \right] =   -63504 \cos^4\theta   + 288 \cos^2\theta  \left(898 \cos^2\theta -9\right)  \left( M/r\right)
\\&&
A^{\varphi \varphi}{}_{2,1} \left[ \theta , M/r \right] =   2160 \left(22 \cos^4\theta +38 \cos^2\theta -75\right)  - 288 \left(382 \cos^4\theta +168 \cos^2\theta -855\right)  \left( M/r\right)
\nonumber \\&& \quad\quad - 240 \cos^2\theta  \left(1104 \cos^2\theta +535\right)  \left( M/r\right)^2
\\&&
A^{\varphi \varphi}{}_{2,2} \left[ \theta , M/r \right] =   96 \left(1035 \cos^4\theta -21 \cos^2\theta -1349\right)  \left( M/r\right)^2+ 16 \cos^2\theta  \left(2069 \cos^2\theta +10189\right)  \left( M/r\right)^3
\\&&
A^{\varphi \varphi}{}_{2,3} \left[ \theta , M/r \right] =   3 \cos^2\theta  \left(5839 \cos^2\theta -16492\right)  \left( M/r\right)^4
\\&&
A^{\varphi \varphi}{}_{3,0} \left[ \theta , M/r \right] =   -63504 \cos^6\theta   - 96 \cos^4\theta  \left(1189 \cos^2\theta +15\right)  \left( M/r\right)
\\&&
A^{\varphi \varphi}{}_{3,1} \left[ \theta , M/r \right] =   -2160 \left(22 \cos^6\theta -132 \cos^4\theta +75 \cos^2\theta +50\right)
\nonumber \\&& \quad\quad + 2880 \cos^2\theta  \left(53 \cos^4\theta -143 \cos^2\theta +129\right)  \left( M/r\right) + 16 \cos^4\theta  \left(6133 \cos^2\theta +645\right)  \left( M/r\right)^2
\\&&
A^{\varphi \varphi}{}_{3,2} \left[ \theta , M/r \right] =   -32 \cos^2\theta  \left(1195 \cos^4\theta -6687 \cos^2\theta +6750\right)  \left( M/r\right)^2
\nonumber \\&& \quad\quad - 8 \cos^4\theta  \left(5047 \cos^2\theta -2061\right)  \left( M/r\right)^3
\\&&
A^{\varphi \varphi}{}_{3,3} \left[ \theta , M/r \right] =   \cos^4\theta  \left(10493 \cos^2\theta -11042\right)  \left( M/r\right)^4
\\&&
A^{\varphi \varphi}{}_{4,0} \left[ \theta , M/r \right] =   40824 \cos^8\theta   + 144 \cos^6\theta  \left(39 \cos^2\theta +10\right)  \left( M/r\right)
\\&&
A^{\varphi \varphi}{}_{4,1} \left[ \theta , M/r \right] =   -1080 \cos^2\theta  \left(41 \cos^6\theta -76 \cos^4\theta -150 \cos^2\theta +200\right)
\nonumber \\&& \quad\quad + 720 \cos^4\theta  \left(41 \cos^4\theta -232 \cos^2\theta +174\right)  \left( M/r\right) - 8 \cos^6\theta  \left(221 \cos^2\theta +158\right)  \left( M/r\right)^2
\\&&
A^{\varphi \varphi}{}_{4,2} \left[ \theta , M/r \right] =   -16 \cos^4\theta  \left(1675 \cos^4\theta -6830 \cos^2\theta +5406\right)  \left( M/r\right)^2
\\&&
A^{\varphi \varphi}{}_{4,3}\left[ \theta , M/r \right] = 0
\\&&
A^{\varphi \varphi}{}_{5,0} \left[ \theta , M/r \right] =   -1512 \cos^{10}\theta
\\&&
A^{\varphi \varphi}{}_{5,1} \left[ \theta , M/r \right] =   1080 \cos^4\theta  \left(\cos^2\theta -2\right) \left(3 \cos^4\theta -50 \cos^2\theta +50\right)
\\&&
A^{\varphi \varphi}{}_{5,2}\left[ \theta , M/r \right] = 0
\\&&
A^{\varphi \varphi}{}_{5,3}\left[ \theta , M/r \right] = 0 .
\end{eqnarray}
\end{subequations}

For the massive Dirac field, the coefficients $A^{t \varphi}{}_{p,q}$ appearing in the expression (\ref{KN_SETtphi}) of $\langle T{}^t{}_\varphi \rangle {}_\mathrm{ren}$ are

\begin{subequations}\label{KN_SETtphiDirac_coeff}
\begin{eqnarray}
&&
A^{t \varphi}{}_{0,0} \left[ \theta , M/r \right] =   576  \left( M/r\right)
\\&&
A^{t \varphi}{}_{0,1} \left[ \theta , M/r \right] =   -10800  + 37296  \left( M/r\right)- 26320  \left( M/r\right)^2
\\&&
A^{t \varphi}{}_{0,2} \left[ \theta , M/r \right] =   -20160  \left( M/r\right)^2+ 27740  \left( M/r\right)^3
\\&&
A^{t \varphi}{}_{0,3} \left[ \theta , M/r \right] =   -7425  \left( M/r\right)^4
\\&&
A^{t \varphi}{}_{1,0} \left[ \theta , M/r \right] =   -144 \left(93 \cos^2\theta -1\right)  \left( M/r\right)
\\&&
A^{t \varphi}{}_{1,1} \left[ \theta , M/r \right] =   10800 \left(\cos^2\theta -6\right)  - 576 \left(39 \cos^2\theta -256\right)  \left( M/r\right)+ 8 \left(3837 \cos^2\theta -7277\right)  \left( M/r\right)^2
\\&&
A^{t \varphi}{}_{1,2} \left[ \theta , M/r \right] =   96 \left(47 \cos^2\theta -817\right)  \left( M/r\right)^2- 4 \left(3360 \cos^2\theta -15373\right)  \left( M/r\right)^3
\\&&
A^{t \varphi}{}_{1,3} \left[ \theta , M/r \right] =  (526 \cos^2\theta -16273) \left( M/r\right)^4
\\&&
A^{t \varphi}{}_{2,0} \left[ \theta , M/r \right] =   144 \cos^2\theta  \left(205 \cos^2\theta -9\right)  \left( M/r\right)
\\&&
A^{t \varphi}{}_{2,1} \left[ \theta , M/r \right] =   21600 \left(2 \cos^4\theta -2 \cos^2\theta -5\right)  - 144 \left(635 \cos^4\theta -774 \cos^2\theta -855\right)  \left( M/r\right)
\nonumber \\&& \quad\quad + 120 \cos^2\theta  \left(19 \cos^2\theta -535\right)  \left( M/r\right)^2
\\&&
A^{t \varphi}{}_{2,2} \left[ \theta , M/r \right] =   48 \left(949 \cos^4\theta -1616 \cos^2\theta -1349\right)  \left( M/r\right)^2- 4 \cos^2\theta  \left(6277 \cos^2\theta -20378\right)  \left( M/r\right)^3
\\&&
A^{t \varphi}{}_{2,3} \left[ \theta , M/r \right] =   \cos^2\theta  \left(8095 \cos^2\theta -24738\right)  \left( M/r\right)^4
\\&&
A^{t \varphi}{}_{3,0} \left[ \theta , M/r \right] =   -144 \cos^4\theta  \left(79 \cos^2\theta +5\right)  \left( M/r\right)
\\&&
A^{t \varphi}{}_{3,1} \left[ \theta , M/r \right] =   10800 \left(\cos^6\theta +9 \cos^4\theta -15 \cos^2\theta -5\right)
\nonumber \\&& \quad\quad - 1440 \cos^2\theta  \left(13 \cos^4\theta +74 \cos^2\theta -129\right)  \left( M/r\right) + 8 \cos^4\theta  \left(2027 \cos^2\theta +645\right)  \left( M/r\right)^2
\\&&
A^{t \varphi}{}_{3,2} \left[ \theta , M/r \right] =   288 \cos^2\theta  \left(50 \cos^4\theta +143 \cos^2\theta -375\right)  \left( M/r\right)^2- 4 \cos^4\theta  \left(3406 \cos^2\theta -2061\right)  \left( M/r\right)^3
\phantom{M/r}
\\&&
A^{t \varphi}{}_{3,3} \left[ \theta , M/r \right] =   \cos^4\theta  \left(4112 \cos^2\theta -5521\right)  \left( M/r\right)^4
\\&&
A^{t \varphi}{}_{4,0} \left[ \theta , M/r \right] =   144 \cos^6\theta  \left(3 \cos^2\theta +5\right)  \left( M/r\right)
\\&&
A^{t \varphi}{}_{4,1} \left[ \theta , M/r \right] =   -10800 \cos^2\theta  \left(\cos^6\theta -6 \cos^4\theta +10\right)  + 720 \cos^4\theta  \left(18 \cos^4\theta -98 \cos^2\theta +87\right)  \left( M/r\right)
\nonumber \\&& \quad\quad - 8 \cos^6\theta  \left(27 \cos^2\theta +79\right)  \left( M/r\right)^2
\\&&
A^{t \varphi}{}_{4,2} \left[ \theta , M/r \right] =   -48 \cos^4\theta  \left(135 \cos^4\theta -840 \cos^2\theta +901\right)  \left( M/r\right)^2
\\&&
A^{t \varphi}{}_{4,3} \left[ \theta , M/r \right] = 0
\\&&
A^{t \varphi}{}_{5,0} \left[ \theta , M/r \right] = 0
\\&&
A^{t \varphi}{}_{5,1} \left[ \theta , M/r \right] =   -10800 \cos^4\theta  \left(\cos^4\theta -5 \cos^2\theta +5\right)
\\&&
A^{t \varphi}{}_{5,2} \left[ \theta , M/r \right] = 0
\\&&
A^{t \varphi}{}_{5,3} \left[ \theta , M/r \right] = 0 .
\end{eqnarray}
\end{subequations}

For the massive Dirac field, the coefficients $A^{\varphi t}{}_{p,q}$ appearing in the expression (\ref{KN_SETphit}) of $\langle T{}^\varphi{}_t \rangle {}_\mathrm{ren}$ are

\begin{subequations}\label{KN_SETphitDirac_coeff}
\begin{eqnarray}
&&
A^{\varphi t}{}_{0,0} \left[ \theta , M/r \right] =   -144  \left( M/r\right)
\\&&
A^{\varphi t}{}_{0,1} \left[ \theta , M/r \right] =   10800  - 50256  \left( M/r\right)+ 58216  \left( M/r\right)^2
\\&&
A^{\varphi t}{}_{0,2} \left[ \theta , M/r \right] =   26640  \left( M/r\right)^2- 61492  \left( M/r\right)^3
\\&&
A^{\varphi t}{}_{0,3} \left[ \theta , M/r \right] =   16273  \left( M/r\right)^4
\\&&
A^{\varphi t}{}_{1,0} \left[ \theta , M/r \right] =   1296 \cos^2\theta   \left( M/r\right)
\\&&
A^{\varphi t}{}_{1,1} \left[ \theta , M/r \right] =   -10800 \left(\cos^2\theta -5\right)  - 432 \left(8 \cos^2\theta +285\right)  \left( M/r\right)+ 64200 \cos^2\theta   \left( M/r\right)^2
\\&&
A^{\varphi t}{}_{1,2} \left[ \theta , M/r \right] =   48 \left(176 \cos^2\theta +1349\right)  \left( M/r\right)^2- 81512 \cos^2\theta   \left( M/r\right)^3
\\&&
A^{\varphi t}{}_{1,3} \left[ \theta , M/r \right] =   24738 \cos^2\theta   \left( M/r\right)^4
\\&&
A^{\varphi t}{}_{2,0} \left[ \theta , M/r \right] =   720 \cos^4\theta   \left( M/r\right)
\\&&
A^{\varphi t}{}_{2,1} \left[ \theta , M/r \right] =   -10800 \left(4 \cos^4\theta -5 \cos^2\theta -5\right) + 720 \cos^2\theta  \left(127 \cos^2\theta -258\right)  \left( M/r\right)
\nonumber \\&& \quad\quad - 5160 \cos^4\theta   \left( M/r\right)^2
\\&&
A^{\varphi t}{}_{2,2} \left[ \theta , M/r \right] =   -48 \cos^2\theta  \left(949 \cos^2\theta -2250\right)  \left( M/r\right)^2- 8244 \cos^4\theta   \left( M/r\right)^3
\\&&
A^{\varphi t}{}_{2,3} \left[ \theta , M/r \right] =   5521 \cos^4\theta   \left( M/r\right)^4
\\&&
A^{\varphi t}{}_{3,0} \left[ \theta , M/r \right] =   -720 \cos^6\theta   \left( M/r\right)
\\&&
A^{\varphi t}{}_{3,1} \left[ \theta , M/r \right] =   -10800 \cos^2\theta  \left(\cos^4\theta +5 \cos^2\theta -10\right)  + 720 \cos^4\theta  \left(62 \cos^2\theta -87\right)  \left( M/r\right)
\nonumber \\&& \quad\quad + 632 \cos^6\theta   \left( M/r\right)^2
\\&&
A^{\varphi t}{}_{3,2} \left[ \theta , M/r \right] =   -48 \cos^4\theta  \left(570 \cos^2\theta -901\right)  \left( M/r\right)^2
\\&&
A^{\varphi t}{}_{3,3} \left[ \theta , M/r \right] = 0
\\&&
A^{\varphi t}{}_{4,0} \left[ \theta , M/r \right] = 0
\\&&
A^{\varphi t}{}_{4,1} \left[ \theta , M/r \right] =   10800 \cos^4\theta  \left(\cos^4\theta -5 \cos^2\theta +5\right)
\\&&
A^{\varphi t}{}_{4,2} \left[ \theta , M/r \right] = 0
\\&&
A^{\varphi t}{}_{4,3} \left[ \theta , M/r \right] = 0 .
\end{eqnarray}
\end{subequations}

For the massive Dirac field, the coefficients $A^{r \theta}{}_{p,q}$ appearing in the expression (\ref{KN_SETrtheta}) of $\langle T{}^r{}_\theta \rangle {}_\mathrm{ren}$ are

\begin{subequations}\label{KN_SETrthetaDirac_coeff}
\begin{eqnarray}
&&
A^{r \theta}{}_{0,0} \left[ \theta , M/r \right] =   -144  + 288  \left( M/r\right)
\\&&
A^{r \theta}{}_{0,1} \left[ \theta , M/r \right] =   279  \left( M/r\right)- 702  \left( M/r\right)^2
\\&&
A^{r \theta}{}_{0,2} \left[ \theta , M/r \right] =   -125  \left( M/r\right)^2+ 529  \left( M/r\right)^3
\\&&
A^{r \theta}{}_{0,3} \left[ \theta , M/r \right] =   -125  \left( M/r\right)^4
\\&&
A^{r \theta}{}_{1,0} \left[ \theta , M/r \right] =   144 \left(7 \cos^2\theta -1\right)  - 2016 \cos^2\theta   \left( M/r\right)
\\&&
A^{r \theta}{}_{1,1} \left[ \theta , M/r \right] =   -9 \left(135 \cos^2\theta -31\right)  \left( M/r\right)+ 3438 \cos^2\theta   \left( M/r\right)^2
\\&&
A^{r \theta}{}_{1,2} \left[ \theta , M/r \right] =  (326 \cos^2\theta -125) \left( M/r\right)^2- 1867 \cos^2\theta   \left( M/r\right)^3
\\&&
A^{r \theta}{}_{1,3} \left[ \theta , M/r \right] =   326 \cos^2\theta   \left( M/r\right)^4
\\&&
A^{r \theta}{}_{2,0} \left[ \theta , M/r \right] =   -1008 \cos^2\theta  \left(\cos^2\theta -1\right)  + 2016 \cos^4\theta   \left( M/r\right)
\\&&
A^{r \theta}{}_{2,1} \left[ \theta , M/r \right] =   45 \cos^2\theta  \left(17 \cos^2\theta -27\right)  \left( M/r\right)- 2538 \cos^4\theta   \left( M/r\right)^2
\\&&
A^{r \theta}{}_{2,2} \left[ \theta , M/r \right] =   -\cos^2\theta  \left(125 \cos^2\theta -326\right)  \left( M/r\right)^2+ 1015 \cos^4\theta   \left( M/r\right)^3
\\&&
A^{r \theta}{}_{2,3} \left[ \theta , M/r \right] =   -125 \cos^4\theta   \left( M/r\right)^4
\\&&
A^{r \theta}{}_{3,0} \left[ \theta , M/r \right] =   144 \cos^4\theta  \left(\cos^2\theta -7\right)  - 288 \cos^6\theta   \left( M/r\right)
\\&&
A^{r \theta}{}_{3,1} \left[ \theta , M/r \right] =   -45 \cos^4\theta  \left(\cos^2\theta -17\right)  \left( M/r\right)+ 234 \cos^6\theta   \left( M/r\right)^2
\\&&
A^{r \theta}{}_{3,2} \left[ \theta , M/r \right] =   -125 \cos^4\theta   \left( M/r\right)^2- 45 \cos^6\theta   \left( M/r\right)^3
\\&&
A^{r \theta}{}_{3,3} \left[ \theta , M/r \right] =   0
\\&&
A^{r \theta}{}_{4,0} \left[ \theta , M/r \right] =   144 \cos^6\theta
\\&&
A^{r \theta}{}_{4,1} \left[ \theta , M/r \right] =   -45 \cos^6\theta   \left( M/r\right)
\\&&
A^{r \theta}{}_{4,2} \left[ \theta , M/r \right] =   0
\\&&
A^{r \theta}{}_{4,3} \left[ \theta , M/r \right] =   0 .
\end{eqnarray}
\end{subequations}

For the massive Dirac field, the coefficients $A^{\theta r}{}_{p,q}$ appearing in the expression (\ref{KN_SETthetar}) of $\langle T{}^\theta{}_r \rangle {}_\mathrm{ren}$ are

\begin{subequations}\label{KN_SETthetarDirac_coeff}
\begin{eqnarray}
&&
A^{\theta r}{}_{0,0} \left[ \theta , M/r \right] =   -144
\\&&
A^{\theta r}{}_{0,1} \left[ \theta , M/r \right] =   279  \left( M/r\right)
\\&&
A^{\theta r}{}_{0,2} \left[ \theta , M/r \right] =   -125  \left( M/r\right)^2
\\&&
A^{\theta r}{}_{1,0} \left[ \theta , M/r \right] =   1008 \cos^2\theta
\\&&
A^{\theta r}{}_{1,1} \left[ \theta , M/r \right] =   -1215 \cos^2\theta   \left( M/r\right)
\\&&
A^{\theta r}{}_{1,2} \left[ \theta , M/r \right] =   326 \cos^2\theta   \left( M/r\right)^2
\\&&
A^{\theta r}{}_{2,0} \left[ \theta , M/r \right] =   -1008 \cos^4\theta
\\&&
A^{\theta r}{}_{2,1} \left[ \theta , M/r \right] =   765 \cos^4\theta   \left( M/r\right)
\\&&
A^{\theta r}{}_{2,2} \left[ \theta , M/r \right] =   -125 \cos^4\theta   \left( M/r\right)^2
\\&&
A^{\theta r}{}_{3,0} \left[ \theta , M/r \right] =   144 \cos^6\theta
\\&&
A^{\theta r}{}_{3,1} \left[ \theta , M/r \right] =   -45 \cos^6\theta   \left( M/r\right)
\\&&
A^{\theta r}{}_{3,2} \left[ \theta , M/r \right] =   0 .
\end{eqnarray}
\end{subequations}

\subsection{Proca field} \label{KN_SETProca_coeff}

For the Proca field, the coefficients $A^{t t}{}_{p,q}$ appearing in the expression (\ref{KN_SETtt}) of $\langle T{}^t{}_t \rangle {}_\mathrm{ren}$ are

\begin{subequations}\label{KN_SETttProca_coeff}
\begin{eqnarray}
&&
A^{t t}{}_{0,0} \left[ \theta , M/r \right] =   6660  - 14664  \left( M/r\right)
\\&&
A^{t t}{}_{0,1} \left[ \theta , M/r \right] =   48600  - 276096  \left( M/r\right)+ 374148  \left( M/r\right)^2
\\&&
A^{t t}{}_{0,2} \left[ \theta , M/r \right] =   167416  \left( M/r\right)^2- 430064  \left( M/r\right)^3
\\&&
A^{t t}{}_{0,3} \left[ \theta , M/r \right] =   124228  \left( M/r\right)^4
\\&&
A^{t t}{}_{1,0} \left[ \theta , M/r \right] =   -179820 \cos^2\theta   + 288 \left(1528 \cos^2\theta +5\right)  \left( M/r\right)
\\&&
A^{t t}{}_{1,1} \left[ \theta , M/r \right] =   -48600 \left(7 \cos^2\theta -12\right)  + 144 \left(10435 \cos^2\theta -13313\right)  \left( M/r\right)
\nonumber \\&& \quad\quad - 48 \left(36771 \cos^2\theta -23602\right)  \left( M/r\right)^2
\\&&
A^{t t}{}_{1,2} \left[ \theta , M/r \right] =   -16 \left(47215 \cos^2\theta -69264\right)  \left( M/r\right)^2+ 16 \left(91476 \cos^2\theta -81229\right)  \left( M/r\right)^3
\\&&
A^{t t}{}_{1,3} \left[ \theta , M/r \right] =   -4 \left(86153 \cos^2\theta -92104\right)  \left( M/r\right)^4
\\&&
A^{t t}{}_{2,0} \left[ \theta , M/r \right] =   279720 \cos^4\theta   - 144 \cos^2\theta  \left(8261 \cos^2\theta +90\right)  \left( M/r\right)
\\&&
A^{t t}{}_{2,1} \left[ \theta , M/r \right] =   -97200 \left(4 \cos^4\theta +6 \cos^2\theta -15\right)  + 144 \left(6824 \cos^4\theta +3147 \cos^2\theta -15845\right)  \left( M/r\right)
\nonumber \\&& \quad\quad + 120 \cos^2\theta  \left(8033 \cos^2\theta +6732\right)  \left( M/r\right)^2
\\&&
A^{t t}{}_{2,2} \left[ \theta , M/r \right] =   -96 \left(8100 \cos^4\theta +290 \cos^2\theta -13289\right)  \left( M/r\right)^2+ 16 \cos^2\theta  \left(13713 \cos^2\theta -73046\right)  \left( M/r\right)^3
\phantom{M/r}
\\&&
A^{t t}{}_{2,3} \left[ \theta , M/r \right] =   -4 \cos^2\theta  \left(51301 \cos^2\theta -97664\right)  \left( M/r\right)^4
\\&&
A^{t t}{}_{3,0} \left[ \theta , M/r \right] =   279720 \cos^6\theta   + 288 \cos^4\theta  \left(1922 \cos^2\theta -25\right)  \left( M/r\right)
\\&&
A^{t t}{}_{3,1} \left[ \theta , M/r \right] =   97200 \left(4 \cos^6\theta -24 \cos^4\theta +15 \cos^2\theta +10\right)
\nonumber \\&& \quad\quad - 240 \cos^2\theta  \left(4403 \cos^4\theta -14871 \cos^2\theta +13386\right)  \left( M/r\right) - 144 \cos^4\theta  \left(2837 \cos^2\theta +1030\right)  \left( M/r\right)^2
\\&&
A^{t t}{}_{3,2} \left[ \theta , M/r \right] =   80 \cos^2\theta  \left(4583 \cos^4\theta -24648 \cos^2\theta +24300\right)  \left( M/r\right)^2
\nonumber \\&& \quad\quad + 16 \cos^4\theta  \left(10654 \cos^2\theta -2425\right)  \left( M/r\right)^3
\\&&
A^{t t}{}_{3,3} \left[ \theta , M/r \right] =   -172 \cos^4\theta  \left(337 \cos^2\theta -376\right)  \left( M/r\right)^4
\\&&
A^{t t}{}_{4,0} \left[ \theta , M/r \right] =   -179820 \cos^8\theta   - 72 \cos^6\theta  \left(493 \cos^2\theta -100\right)  \left( M/r\right)
\\&&
A^{t t}{}_{4,1} \left[ \theta , M/r \right] =   48600 \cos^2\theta  \left(7 \cos^6\theta -12 \cos^4\theta -30 \cos^2\theta +40\right)
\nonumber \\&& \quad\quad - 720 \cos^4\theta  \left(362 \cos^4\theta -1665 \cos^2\theta +1293\right)  \left( M/r\right) + 12 \cos^6\theta  \left(675 \cos^2\theta +584\right)  \left( M/r\right)^2
\\&&
A^{t t}{}_{4,2} \left[ \theta , M/r \right] =   8 \cos^4\theta  \left(27673 \cos^4\theta -104472 \cos^2\theta +83532\right)  \left( M/r\right)^2
\\&&
A^{t t}{}_{4,3} \left[ \theta , M/r \right] = 0
\\&&
A^{t t}{}_{5,0} \left[ \theta , M/r \right] =   6660 \cos^{10}\theta
\\&&
A^{t t}{}_{5,1} \left[ \theta , M/r \right] =   -48600 \cos^4\theta  \left(\cos^2\theta -2\right) \left(\cos^4\theta -10 \cos^2\theta +10\right)
\\&&
A^{t t}{}_{5,2} \left[ \theta , M/r \right] =  0
\\&&
A^{t t}{}_{5,3} \left[ \theta , M/r \right] =  0 .
\end{eqnarray}
\end{subequations}

For the Proca field, the coefficients $A^{r r}{}_{p,q}$ appearing in the expression (\ref{KN_SETrr}) of $\langle T{}^r{}_r \rangle {}_\mathrm{ren}$ are

\begin{subequations}\label{KN_SETrrProca_coeff}
\begin{eqnarray}
&&
A^{r r}{}_{0,0} \left[ \theta , M/r \right] =   -2772  + 4200  \left( M/r\right)
\\&&
A^{r r}{}_{0,1} \left[ \theta , M/r \right] =   9720  - 41792  \left( M/r\right)+ 51628  \left( M/r\right)^2
\\&&
A^{r r}{}_{0,2} \left[ \theta , M/r \right] =   25768  \left( M/r\right)^2- 67984  \left( M/r\right)^3
\\&&
A^{r r}{}_{0,3} \left[ \theta , M/r \right] =   21460  \left( M/r\right)^4
\\&&
A^{r r}{}_{1,0} \left[ \theta , M/r \right] =   2772 \left(31 \cos^2\theta -4\right)  - 86688 \cos^2\theta   \left( M/r\right)
\\&&
A^{r r}{}_{1,1} \left[ \theta , M/r \right] =   -9720 \left(3 \cos^2\theta -8\right)  - 288 \left(301 \cos^2\theta +624\right)  \left( M/r\right)+ 271088 \cos^2\theta   \left( M/r\right)^2
\\&&
A^{r r}{}_{1,2} \left[ \theta , M/r \right] =   8 \left(3633 \cos^2\theta +13213\right)  \left( M/r\right)^2- 236144 \cos^2\theta   \left( M/r\right)^3
\\&&
A^{r r}{}_{1,3} \left[ \theta , M/r \right] =   64076 \cos^2\theta   \left( M/r\right)^4
\\&&
A^{r r}{}_{2,0} \left[ \theta , M/r \right] =   -38808 \cos^2\theta  \left(11 \cos^2\theta -8\right)  + 117936 \cos^4\theta   \left( M/r\right)
\\&&
A^{r r}{}_{2,1} \left[ \theta , M/r \right] =   -19440 \left(6 \cos^4\theta -6 \cos^2\theta -5\right)  + 96 \cos^2\theta  \left(7673 \cos^2\theta -8766\right)  \left( M/r\right)
\nonumber \\&& \quad\quad - 76280 \cos^4\theta   \left( M/r\right)^2
\\&&
A^{r r}{}_{2,2} \left[ \theta , M/r \right] =   -8 \cos^2\theta  \left(34243 \cos^2\theta -48823\right)  \left( M/r\right)^2- 34096 \cos^4\theta   \left( M/r\right)^3
\\&&
A^{r r}{}_{2,3} \left[ \theta , M/r \right] =   24796 \cos^4\theta   \left( M/r\right)^4
\\&&
A^{r r}{}_{3,0} \left[ \theta , M/r \right] =   38808 \cos^4\theta  \left(17 \cos^2\theta -20\right)  + 18144 \cos^6\theta   \left( M/r\right)
\\&&
A^{r r}{}_{3,1} \left[ \theta , M/r \right] =   -19440 \cos^2\theta  \left(4 \cos^4\theta +6 \cos^2\theta -15\right)  - 160 \cos^4\theta  \left(2963 \cos^2\theta -3384\right)  \left( M/r\right)
\nonumber \\&& \quad\quad - 18192 \cos^6\theta   \left( M/r\right)^2
\\&&
A^{r r}{}_{3,2} \left[ \theta , M/r \right] =   8 \cos^4\theta  \left(5571 \cos^2\theta -2977\right)  \left( M/r\right)^2- 1872 \cos^6\theta   \left( M/r\right)^3
\\&&
A^{r r}{}_{3,3} \left[ \theta , M/r \right] =   4836 \cos^6\theta   \left( M/r\right)^4
\\&&
A^{r r}{}_{4,0} \left[ \theta , M/r \right] =   -2772 \cos^6\theta  \left(85 \cos^2\theta -112\right)  - 9432 \cos^8\theta   \left( M/r\right)
\\&&
A^{r r}{}_{4,1} \left[ \theta , M/r \right] =   9720 \cos^4\theta  \left(3 \cos^4\theta -28 \cos^2\theta +30\right)  + 480 \cos^6\theta  \left(341 \cos^2\theta -450\right)  \left( M/r\right)
\nonumber \\&& \quad\quad + 5676 \cos^8\theta   \left( M/r\right)^2
\\&&
A^{r r}{}_{4,2} \left[ \theta , M/r \right] =   -8 \cos^6\theta  \left(4126 \cos^2\theta -5765\right)  \left( M/r\right)^2
\\&&
A^{r r}{}_{4,3} \left[ \theta , M/r \right] = 0
\\&&
A^{r r}{}_{5,0} \left[ \theta , M/r \right] =   2772 \cos^8\theta  \left(3 \cos^2\theta -4\right)
\\&&
A^{r r}{}_{5,1} \left[ \theta , M/r \right] =   9720 \cos^6\theta  \left(3 \cos^4\theta -12 \cos^2\theta +10\right)
\\&&
A^{r r}{}_{5,2} \left[ \theta , M/r \right] = 0
\\&&
A^{r r}{}_{5,3} \left[ \theta , M/r \right] = 0 .
\end{eqnarray}
\end{subequations}

For the Proca field, the coefficients $A^{\theta \theta}{}_{p,q}$ appearing in the expression (\ref{KN_SETthetatheta}) of $\langle T{}^\theta{}_\theta \rangle {}_\mathrm{ren}$ are

\begin{subequations}\label{KN_SETthetathetaProca_coeff}
\begin{eqnarray}
&&
A^{\theta \theta}{}_{0,0} \left[ \theta , M/r \right] =   8316  - 19416  \left( M/r\right)
\\&&
A^{\theta \theta}{}_{0,1} \left[ \theta , M/r \right] =   -29160  + 126832  \left( M/r\right)- 123524  \left( M/r\right)^2
\\&&
A^{\theta \theta}{}_{0,2} \left[ \theta , M/r \right] =   -83632  \left( M/r\right)^2+ 176272  \left( M/r\right)^3
\\&&
A^{\theta \theta}{}_{0,3} \left[ \theta , M/r \right] =   -55916  \left( M/r\right)^4
\\&&
A^{\theta \theta}{}_{1,0} \left[ \theta , M/r \right] =   -2772 \left(85 \cos^2\theta -4\right)  + 545760 \cos^2\theta   \left( M/r\right)
\\&&
A^{\theta \theta}{}_{1,1} \left[ \theta , M/r \right] =   -29160 \left(\cos^2\theta +4\right)  + 4176 \left(143 \cos^2\theta +59\right)  \left( M/r\right)- 1519024 \cos^2\theta   \left( M/r\right)^2
\\&&
A^{\theta \theta}{}_{1,2} \left[ \theta , M/r \right] =   -8 \left(39591 \cos^2\theta +18535\right)  \left( M/r\right)^2+ 1292144 \cos^2\theta   \left( M/r\right)^3
\\&&
A^{\theta \theta}{}_{1,3} \left[ \theta , M/r \right] =   -340212 \cos^2\theta   \left( M/r\right)^4
\\&&
A^{\theta \theta}{}_{2,0} \left[ \theta , M/r \right] =   38808 \cos^2\theta  \left(17 \cos^2\theta -8\right)  - 1414224 \cos^4\theta   \left( M/r\right)
\\&&
A^{\theta \theta}{}_{2,1} \left[ \theta , M/r \right] =   19440 \left(4 \cos^4\theta -14 \cos^2\theta -5\right)  - 48 \cos^2\theta  \left(17171 \cos^2\theta -24471\right)  \left( M/r\right)
\nonumber \\&& \quad\quad + 2057800 \cos^4\theta   \left( M/r\right)^2
\\&&
A^{\theta \theta}{}_{2,2} \left[ \theta , M/r \right] =   8 \cos^2\theta  \left(19663 \cos^2\theta -75877\right)  \left( M/r\right)^2- 766160 \cos^4\theta   \left( M/r\right)^3
\\&&
A^{\theta \theta}{}_{2,3} \left[ \theta , M/r \right] =   34908 \cos^4\theta   \left( M/r\right)^4
\\&&
A^{\theta \theta}{}_{3,0} \left[ \theta , M/r \right] =   -38808 \cos^4\theta  \left(11 \cos^2\theta -20\right)  + 639072 \cos^6\theta   \left( M/r\right)
\\&&
A^{\theta \theta}{}_{3,1} \left[ \theta , M/r \right] =   58320 \cos^2\theta  \left(2 \cos^4\theta -2 \cos^2\theta -5\right)  - 80 \cos^4\theta  \left(67 \cos^2\theta +3159\right)  \left( M/r\right)
\nonumber \\&& \quad\quad - 522864 \cos^6\theta   \left( M/r\right)^2
\\&&
A^{\theta \theta}{}_{3,2} \left[ \theta , M/r \right] =   8 \cos^4\theta  \left(10947 \cos^2\theta -24077\right)  \left( M/r\right)^2+ 110928 \cos^6\theta   \left( M/r\right)^3
\\&&
A^{\theta \theta}{}_{3,3} \left[ \theta , M/r \right] =   -12956 \cos^6\theta   \left( M/r\right)^4
\\&&
A^{\theta \theta}{}_{4,0} \left[ \theta , M/r \right] =   2772 \cos^6\theta  \left(31 \cos^2\theta -112\right)  - 38808 \cos^8\theta   \left( M/r\right)
\\&&
A^{\theta \theta}{}_{4,1} \left[ \theta , M/r \right] =   29160 \cos^4\theta  \left(\cos^4\theta +4 \cos^2\theta -10\right)  - 240 \cos^6\theta  \left(542 \cos^2\theta -993\right)  \left( M/r\right)
\nonumber \\&& \quad\quad + 9756 \cos^8\theta   \left( M/r\right)^2
\\&&
A^{\theta \theta}{}_{4,2} \left[ \theta , M/r \right] =   8 \cos^6\theta  \left(6499 \cos^2\theta -11087\right)  \left( M/r\right)^2
\\&&
A^{\theta \theta}{}_{4,3} \left[ \theta , M/r \right] = 0
\\&&
A^{\theta \theta}{}_{5,0} \left[ \theta , M/r \right] =   -2772 \cos^8\theta  \left(\cos^2\theta -4\right)
\\&&
A^{\theta \theta}{}_{5,1} \left[ \theta , M/r \right] =   -9720 \cos^6\theta  \left(\cos^4\theta -8 \cos^2\theta +10\right)
\\&&
A^{\theta \theta}{}_{5,2} \left[ \theta , M/r \right] = 0
\\&&
A^{\theta \theta}{}_{5,3} \left[ \theta , M/r \right] = 0 .
\end{eqnarray}
\end{subequations}

For the Proca field, the coefficients $A^{\varphi \varphi}{}_{p,q}$ appearing in the expression (\ref{KN_SETphiphi}) of $\langle T{}^\varphi{}_\varphi \rangle {}_\mathrm{ren}$ are

\begin{subequations}\label{KN_SETphiphiProca_coeff}
\begin{eqnarray}
&&
A^{\varphi \varphi}{}_{0,0} \left[ \theta , M/r \right] =   8316  - 19416  \left( M/r\right)
\\&&
A^{\varphi \varphi}{}_{0,1} \left[ \theta , M/r \right] =   -29160  + 126832  \left( M/r\right)- 123524  \left( M/r\right)^2
\\&&
A^{\varphi \varphi}{}_{0,2} \left[ \theta , M/r \right] =   -83632  \left( M/r\right)^2+ 176272  \left( M/r\right)^3
\\&&
A^{\varphi \varphi}{}_{0,3} \left[ \theta , M/r \right] =   -55916  \left( M/r\right)^4
\\&&
A^{\varphi \varphi}{}_{1,0} \left[ \theta , M/r \right] =   -224532 \cos^2\theta   + 1440 \left(380 \cos^2\theta -1\right)  \left( M/r\right)
\\&&
A^{\varphi \varphi}{}_{1,1} \left[ \theta , M/r \right] =   9720 \left(41 \cos^2\theta -56\right)  - 8352 \left(115 \cos^2\theta -216\right)  \left( M/r\right)
\nonumber \\&& \quad\quad - 16 \left(24133 \cos^2\theta +70806\right)  \left( M/r\right)^2
\\&&
A^{\varphi \varphi}{}_{1,2} \left[ \theta , M/r \right] =   16 \left(36322 \cos^2\theta -65385\right)  \left( M/r\right)^2- 16 \left(470 \cos^2\theta -81229\right)  \left( M/r\right)^3
\\&&
A^{\varphi \varphi}{}_{1,3} \left[ \theta , M/r \right] =   4 \left(7051 \cos^2\theta -92104\right)  \left( M/r\right)^4
\\&&
A^{\varphi \varphi}{}_{2,0} \left[ \theta , M/r \right] =   349272 \cos^4\theta   - 144 \cos^2\theta  \left(9911 \cos^2\theta -90\right)  \left( M/r\right)
\\&&
A^{\varphi \varphi}{}_{2,1} \left[ \theta , M/r \right] =   19440 \left(22 \cos^4\theta +38 \cos^2\theta -75\right)  - 48 \left(32549 \cos^4\theta +7686 \cos^2\theta -47535\right)  \left( M/r\right)
\nonumber \\&& \quad\quad + 40 \cos^2\theta  \left(71641 \cos^2\theta -20196\right)  \left( M/r\right)^2
\\&&
A^{\varphi \varphi}{}_{2,2} \left[ \theta , M/r \right] =   48 \left(18630 \cos^4\theta -1421 \cos^2\theta -26578\right)  \left( M/r\right)^2
\nonumber \\&& \quad\quad - 16 \cos^2\theta  \left(120931 \cos^2\theta -73046\right)  \left( M/r\right)^3
\\&&
A^{\varphi \varphi}{}_{2,3} \left[ \theta , M/r \right] =   4 \cos^2\theta  \left(106391 \cos^2\theta -97664\right)  \left( M/r\right)^4
\\&&
A^{\varphi \varphi}{}_{3,0} \left[ \theta , M/r \right] =   349272 \cos^6\theta   + 288 \cos^4\theta  \left(2194 \cos^2\theta +25\right)  \left( M/r\right)
\\&&
A^{\varphi \varphi}{}_{3,1} \left[ \theta , M/r \right] =   -19440 \left(22 \cos^6\theta -132 \cos^4\theta +75 \cos^2\theta +50\right)
\nonumber \\&& \quad\quad + 160 \cos^2\theta  \left(970 \cos^4\theta -22662 \cos^2\theta +20079\right)  \left( M/r\right) - 144 \cos^4\theta  \left(4661 \cos^2\theta -1030\right)  \left( M/r\right)^2
\\&&
A^{\varphi \varphi}{}_{3,2} \left[ \theta , M/r \right] =   -16 \cos^2\theta  \left(2302 \cos^4\theta -117237 \cos^2\theta +121500\right)  \left( M/r\right)^2
\nonumber \\&& \quad\quad + 16 \cos^4\theta  \left(4508 \cos^2\theta +2425\right)  \left( M/r\right)^3
\\&&
A^{\varphi \varphi}{}_{3,3} \left[ \theta , M/r \right] =   4 \cos^4\theta  \left(12929 \cos^2\theta -16168\right)  \left( M/r\right)^4
\\&&
A^{\varphi \varphi}{}_{4,0} \left[ \theta , M/r \right] =   -224532 \cos^8\theta   - 72 \cos^6\theta  \left(439 \cos^2\theta +100\right)  \left( M/r\right)
\\&&
A^{\varphi \varphi}{}_{4,1} \left[ \theta , M/r \right] =   -9720 \cos^2\theta  \left(41 \cos^6\theta -76 \cos^4\theta -150 \cos^2\theta +200\right)
\nonumber \\&& \quad\quad + 240 \cos^4\theta  \left(2626 \cos^4\theta -6054 \cos^2\theta +3879\right)  \left( M/r\right) + 12 \cos^6\theta  \left(1397 \cos^2\theta -584\right)  \left( M/r\right)^2
\\&&
A^{\varphi \varphi}{}_{4,2} \left[ \theta , M/r \right] =   -16 \cos^4\theta  \left(16643 \cos^4\theta -56115 \cos^2\theta +41766\right)  \left( M/r\right)^2
\\&&
A^{\varphi \varphi}{}_{4,3} \left[ \theta , M/r \right] = 0
\\&&
A^{\varphi \varphi}{}_{5,0} \left[ \theta , M/r \right] =   8316 \cos^{10}\theta
\\&&
A^{\varphi \varphi}{}_{5,1} \left[ \theta , M/r \right] =   9720 \cos^4\theta  \left(\cos^2\theta -2\right) \left(3 \cos^4\theta -50 \cos^2\theta +50\right)
\\&&
A^{\varphi \varphi}{}_{5,2} \left[ \theta , M/r \right] =  0
\\&&
A^{\varphi \varphi}{}_{5,3} \left[ \theta , M/r \right] =  0 .
\end{eqnarray}
\end{subequations}

For the Proca field, the coefficients $A^{t \varphi}{}_{p,q}$ appearing in the expression (\ref{KN_SETtphi}) of $\langle T{}^t{}_\varphi \rangle {}_\mathrm{ren}$ are

\begin{subequations}\label{KN_SETtphiProca_coeff}
\begin{eqnarray}
&&
A^{t \varphi}{}_{0,0} \left[ \theta , M/r \right] =   -2376  \left( M/r\right)
\\&&
A^{t \varphi}{}_{0,1} \left[ \theta , M/r \right] =   -97200  + 359856  \left( M/r\right)- 248836  \left( M/r\right)^2
\\&&
A^{t \varphi}{}_{0,2} \left[ \theta , M/r \right] =   -219660  \left( M/r\right)^2+ 303168  \left( M/r\right)^3
\\&&
A^{t \varphi}{}_{0,3} \left[ \theta , M/r \right] =   -90072  \left( M/r\right)^4
\\&&
A^{t \varphi}{}_{1,0} \left[ \theta , M/r \right] =   144 \left(367 \cos^2\theta -5\right)  \left( M/r\right)
\\&&
A^{t \varphi}{}_{1,1} \left[ \theta , M/r \right] =   97200 \left(\cos^2\theta -6\right)  - 72 \left(4467 \cos^2\theta -19223\right)  \left( M/r\right)+ 16 \left(7687 \cos^2\theta -35403\right)  \left( M/r\right)^2
\phantom{M/r}
\\&&
A^{t \varphi}{}_{1,2} \left[ \theta , M/r \right] =   12 \left(8221 \cos^2\theta -66601\right)  \left( M/r\right)^2- 8 \left(10717 \cos^2\theta -81229\right)  \left( M/r\right)^3
\\&&
A^{t \varphi}{}_{1,3} \left[ \theta , M/r \right] =   8 \left(275 \cos^2\theta -23026\right)  \left( M/r\right)^4
\\&&
A^{t \varphi}{}_{2,0} \left[ \theta , M/r \right] =   -2160 \cos^2\theta  \left(52 \cos^2\theta -3\right)  \left( M/r\right)
\\&&
A^{t \varphi}{}_{2,1} \left[ \theta , M/r \right] =   194400 \left(2 \cos^4\theta -2 \cos^2\theta -5\right)  - 72 \left(12200 \cos^4\theta -11363 \cos^2\theta -15845\right)  \left( M/r\right)
\nonumber \\&& \quad\quad + 440 \cos^2\theta  \left(1243 \cos^2\theta -918\right)  \left( M/r\right)^2
\\&&
A^{t \varphi}{}_{2,2} \left[ \theta , M/r \right] =   12 \left(39001 \cos^4\theta -53339 \cos^2\theta -53156\right)  \left( M/r\right)^2
\nonumber \\&& \quad\quad - 16 \cos^2\theta  \left(30799 \cos^2\theta -36523\right)  \left( M/r\right)^3
\\&&
A^{t \varphi}{}_{2,3} \left[ \theta , M/r \right] =   8 \cos^2\theta  \left(15007 \cos^2\theta -24416\right)  \left( M/r\right)^4
\\&&
A^{t \varphi}{}_{3,0} \left[ \theta , M/r \right] =   144 \cos^4\theta  \left(297 \cos^2\theta +25\right)  \left( M/r\right)
\\&&
A^{t \varphi}{}_{3,1} \left[ \theta , M/r \right] =   97200 \left(\cos^6\theta +9 \cos^4\theta -15 \cos^2\theta -5\right)
\nonumber \\&& \quad\quad - 360 \cos^2\theta  \left(223 \cos^4\theta +3091 \cos^2\theta -4462\right)  \left( M/r\right) - 144 \cos^4\theta  \left(397 \cos^2\theta -515\right)  \left( M/r\right)^2
\\&&
A^{t \varphi}{}_{3,2} \left[ \theta , M/r \right] =   12 \cos^2\theta  \left(7615 \cos^4\theta +40317 \cos^2\theta -81000\right)  \left( M/r\right)^2
\nonumber \\&& \quad\quad - 8 \cos^4\theta  \left(3721 \cos^2\theta -2425\right)  \left( M/r\right)^3
\\&&
A^{t \varphi}{}_{3,3} \left[ \theta , M/r \right] =   8 \cos^4\theta  \left(2813 \cos^2\theta -4042\right)  \left( M/r\right)^4
\\&&
A^{t \varphi}{}_{4,0} \left[ \theta , M/r \right] =   -72 \cos^6\theta  \left(23 \cos^2\theta +50\right)  \left( M/r\right)
\\&&
A^{t \varphi}{}_{4,1} \left[ \theta , M/r \right] =   -97200 \cos^2\theta  \left(\cos^6\theta -6 \cos^4\theta +10\right)
\nonumber \\&& \quad\quad + 360 \cos^4\theta  \left(324 \cos^4\theta -1519 \cos^2\theta +1293\right)  \left( M/r\right) + 12 \cos^6\theta  \left(69 \cos^2\theta -292\right)  \left( M/r\right)^2
\\&&
A^{t \varphi}{}_{4,2} \left[ \theta , M/r \right] =   -12 \cos^4\theta  \left(4860 \cos^4\theta -27055 \cos^2\theta +27844\right)  \left( M/r\right)^2
\\&&
A^{t \varphi}{}_{4,3} \left[ \theta , M/r \right] = 0
\\&&
A^{t \varphi}{}_{5,0} \left[ \theta , M/r \right] = 0
\\&&
A^{t \varphi}{}_{5,1} \left[ \theta , M/r \right] =   -97200 \cos^4\theta  \left(\cos^4\theta -5 \cos^2\theta +5\right)
\\&&
A^{t \varphi}{}_{5,2} \left[ \theta , M/r \right] = 0
\\&&
A^{t \varphi}{}_{5,3} \left[ \theta , M/r \right] = 0 .
\end{eqnarray}
\end{subequations}

For the Proca field, the coefficients $A^{\varphi t}{}_{p,q}$ appearing in the expression (\ref{KN_SETphit}) of $\langle T{}^\varphi{}_t \rangle {}_\mathrm{ren}$ are

\begin{subequations}\label{KN_SETphitProca_coeff}
\begin{eqnarray}
&&
A^{\varphi t}{}_{0,0} \left[ \theta , M/r \right] =   720  \left( M/r\right)
\\&&
A^{\varphi t}{}_{0,1} \left[ \theta , M/r \right] =   97200  - 476496  \left( M/r\right)+ 566448  \left( M/r\right)^2
\\&&
A^{\varphi t}{}_{0,2} \left[ \theta , M/r \right] =   277980  \left( M/r\right)^2- 649832  \left( M/r\right)^3
\\&&
A^{\varphi t}{}_{0,3} \left[ \theta , M/r \right] =   184208  \left( M/r\right)^4
\\&&
A^{\varphi t}{}_{1,0} \left[ \theta , M/r \right] =   -6480 \cos^2\theta   \left( M/r\right)
\\&&
A^{\varphi t}{}_{1,1} \left[ \theta , M/r \right] =   -97200 \left(\cos^2\theta -5\right)  + 72 \left(1227 \cos^2\theta -15845\right)  \left( M/r\right)+ 403920 \cos^2\theta   \left( M/r\right)^2
\\&&
A^{\varphi t}{}_{1,2} \left[ \theta , M/r \right] =   12 \left(1499 \cos^2\theta +53156\right)  \left( M/r\right)^2- 584368 \cos^2\theta   \left( M/r\right)^3
\\&&
A^{\varphi t}{}_{1,3} \left[ \theta , M/r \right] =   195328 \cos^2\theta   \left( M/r\right)^4
\\&&
A^{\varphi t}{}_{2,0} \left[ \theta , M/r \right] =   -3600 \cos^4\theta   \left( M/r\right)
\\&&
A^{\varphi t}{}_{2,1} \left[ \theta , M/r \right] =   -97200 \left(4 \cos^4\theta -5 \cos^2\theta -5\right)  + 720 \cos^2\theta  \left(1220 \cos^2\theta -2231\right)  \left( M/r\right)
\nonumber \\&& \quad\quad - 74160 \cos^4\theta   \left( M/r\right)^2
\\&&
A^{\varphi t}{}_{2,2} \left[ \theta , M/r \right] =   -12 \cos^2\theta  \left(39001 \cos^2\theta -81000\right)  \left( M/r\right)^2- 19400 \cos^4\theta   \left( M/r\right)^3
\\&&
A^{\varphi t}{}_{2,3} \left[ \theta , M/r \right] =   32336 \cos^4\theta   \left( M/r\right)^4
\\&&
A^{\varphi t}{}_{3,0} \left[ \theta , M/r \right] =   3600 \cos^6\theta   \left( M/r\right)
\\&&
A^{\varphi t}{}_{3,1} \left[ \theta , M/r \right] =   -97200 \cos^2\theta  \left(\cos^4\theta +5 \cos^2\theta -10\right)  + 360 \cos^4\theta  \left(871 \cos^2\theta -1293\right)  \left( M/r\right)
\nonumber \\&& \quad\quad + 3504 \cos^6\theta   \left( M/r\right)^2
\\&&
A^{\varphi t}{}_{3,2} \left[ \theta , M/r \right] =   -12 \cos^4\theta  \left(17335 \cos^2\theta -27844\right)  \left( M/r\right)^2
\\&&
A^{\varphi t}{}_{3,3} \left[ \theta , M/r \right] = 0
\\&&
A^{\varphi t}{}_{4,0} \left[ \theta , M/r \right] = 0
\\&&
A^{\varphi t}{}_{4,1} \left[ \theta , M/r \right] =   97200 \cos^4\theta  \left(\cos^4\theta -5 \cos^2\theta +5\right)
\\&&
A^{\varphi t}{}_{4,2} \left[ \theta , M/r \right] = 0
\\&&
A^{\varphi t}{}_{4,3} \left[ \theta , M/r \right] = 0 .
\end{eqnarray}
\end{subequations}

For the Proca field, the coefficients $A^{r \theta}{}_{p,q}$ appearing in the expression (\ref{KN_SETrtheta}) of $\langle T{}^r{}_\theta \rangle {}_\mathrm{ren}$ are

\begin{subequations}\label{KN_SETrthetaProca_coeff}
\begin{eqnarray}
&&
A^{r \theta}{}_{0,0} \left[ \theta , M/r \right] =   792  - 1584  \left( M/r\right)
\\&&
A^{r \theta}{}_{0,1} \left[ \theta , M/r \right] =   -972  \left( M/r\right)+ 2736  \left( M/r\right)^2
\\&&
A^{r \theta}{}_{0,2} \left[ \theta , M/r \right] =   250  \left( M/r\right)^2- 1472  \left( M/r\right)^3
\\&&
A^{r \theta}{}_{0,3} \left[ \theta , M/r \right] =   250  \left( M/r\right)^4
\\&&
A^{r \theta}{}_{1,0} \left[ \theta , M/r \right] =   -792 \left(7 \cos^2\theta -1\right)  + 11088 \cos^2\theta   \left( M/r\right)
\\&&
A^{r \theta}{}_{1,1} \left[ \theta , M/r \right] =   12 \left(635 \cos^2\theta -81\right)  \left( M/r\right)- 20784 \cos^2\theta   \left( M/r\right)^2
\\&&
A^{r \theta}{}_{1,2} \left[ \theta , M/r \right] =   -2 \left(1334 \cos^2\theta -125\right)  \left( M/r\right)^2+ 12956 \cos^2\theta   \left( M/r\right)^3
\\&&
A^{r \theta}{}_{1,3} \left[ \theta , M/r \right] =   -2668 \cos^2\theta   \left( M/r\right)^4
\\&&
A^{r \theta}{}_{2,0} \left[ \theta , M/r \right] =   5544 (\cos\theta -1) \cos^2\theta  (\cos\theta +1)  - 11088 \cos^4\theta   \left( M/r\right)
\\&&
A^{r \theta}{}_{2,1} \left[ \theta , M/r \right] =   -60 \cos^2\theta  \left(67 \cos^2\theta -127\right)  \left( M/r\right)+ 13584 \cos^4\theta   \left( M/r\right)^2
\\&&
A^{r \theta}{}_{2,2} \left[ \theta , M/r \right] =   2 \cos^2\theta  \left(125 \cos^2\theta -1334\right)  \left( M/r\right)^2- 4520 \cos^4\theta   \left( M/r\right)^3
\\&&
A^{r \theta}{}_{2,3} \left[ \theta , M/r \right] =   250 \cos^4\theta   \left( M/r\right)^4
\\&&
A^{r \theta}{}_{3,0} \left[ \theta , M/r \right] =   -792 \cos^4\theta  \left(\cos^2\theta -7\right)  + 1584 \cos^6\theta   \left( M/r\right)
\\&&
A^{r \theta}{}_{3,1} \left[ \theta , M/r \right] =   60 \cos^4\theta  \left(\cos^2\theta -67\right)  \left( M/r\right)- 912 \cos^6\theta   \left( M/r\right)^2
\\&&
A^{r \theta}{}_{3,2} \left[ \theta , M/r \right] =   250 \cos^4\theta   \left( M/r\right)^2+ 60 \cos^6\theta   \left( M/r\right)^3
\\&&
A^{r \theta}{}_{3,3} \left[ \theta , M/r \right] = 0
\\&&
A^{r \theta}{}_{4,0} \left[ \theta , M/r \right] =   -792 \cos^6\theta
\\&&
A^{r \theta}{}_{4,1} \left[ \theta , M/r \right] =   60 \cos^6\theta   \left( M/r\right)
\\&&
A^{r \theta}{}_{4,2} \left[ \theta , M/r \right] = 0
\\&&
A^{r \theta}{}_{4,3} \left[ \theta , M/r \right] = 0 .
\end{eqnarray}
\end{subequations}

For the Proca field, the coefficients $A^{\theta r}{}_{p,q}$ appearing in the expression (\ref{KN_SETthetar}) of $\langle T{}^\theta{}_r \rangle {}_\mathrm{ren}$ are

\begin{subequations}\label{KN_SETthetarProca_coeff}
\begin{eqnarray}
&&
A^{\theta r}{}_{0,0} \left[ \theta , M/r \right] =   792
\\&&
A^{\theta r}{}_{0,1} \left[ \theta , M/r \right] =   -972  \left( M/r\right)
\\&&
A^{\theta r}{}_{0,2} \left[ \theta , M/r \right] =   250  \left( M/r\right)^2
\\&&
A^{\theta r}{}_{1,0} \left[ \theta , M/r \right] =   -5544 \cos^2\theta
\\&&
A^{\theta r}{}_{1,1} \left[ \theta , M/r \right] =   7620 \cos^2\theta   \left( M/r\right)
\\&&
A^{\theta r}{}_{1,2} \left[ \theta , M/r \right] =   -2668 \cos^2\theta   \left( M/r\right)^2
\\&&
A^{\theta r}{}_{2,0} \left[ \theta , M/r \right] =   5544 \cos^4\theta
\\&&
A^{\theta r}{}_{2,1} \left[ \theta , M/r \right] =   -4020 \cos^4\theta   \left( M/r\right)
\\&&
A^{\theta r}{}_{2,2} \left[ \theta , M/r \right] =   250 \cos^4\theta   \left( M/r\right)^2
\\&&
A^{\theta r}{}_{3,0} \left[ \theta , M/r \right] =   -792 \cos^6\theta
\\&&
A^{\theta r}{}_{3,1} \left[ \theta , M/r \right] =   60 \cos^6\theta   \left( M/r\right)
\\&&
A^{\theta r}{}_{3,2} \left[ \theta , M/r \right] = 0 .
\end{eqnarray}
\end{subequations}

\end{widetext}

\bibliography{RSET_in_KN}

\end{document}